\documentclass[11pt,a4paper]{article}
\usepackage{lscape}
\usepackage{mathtools}

    \usepackage{subcaption}
\usepackage[utf8]{inputenc}		
\usepackage[T1]{fontenc}
\usepackage{multirow}
\usepackage{graphicx}
\usepackage{natbib}
\usepackage[dvipsnames]{xcolor}
\usepackage{booktabs}
\usepackage{rotating}
\usepackage{multirow}
\usepackage{graphicx}
\newcommand{\biblist}{\begin{list}{}
		{\listparindent 0.0cm \leftmargin 0.50cm \itemindent -0.50 cm
			\labelwidth 0 cm \labelsep 0.50 cm
			\usecounter{list}}\clubpenalty4000\widowpenalty4000}
	\newcommand{\ebiblist}{\end{list}}

\usepackage{amsmath,amsthm,amsfonts,amssymb,amscd,amsthm,xspace, dsfont}
\theoremstyle{plain}
\allowdisplaybreaks

\usepackage{latexsym}
\usepackage{amsmath}
\usepackage{amsfonts,amssymb}
\usepackage{multirow}
\usepackage{enumerate}

\newtheorem*{theorem*}{Theorem}

\newtheorem{remark}{Remark}[section]
\newtheorem{proposition}{Proposition}[section]
\newtheorem*{proposition*}{Proposition}
\newtheorem{result}{Result}[section]
\newtheorem*{result*}{Result}

\newtheorem*{lemma*}{Lemma}

\newcommand{\bx}{\mathbf{x}}

\newcommand{\V}{{\mathbb V}}

\usepackage{hyperref}
\hypersetup{
	colorlinks=true,
	linkcolor=blue,
	filecolor=magenta,
	urlcolor=blue,
	pdftitle={Sharelatex Example},
	bookmarks=true,
	citecolor=blue,
}

\DeclareMathOperator*{\argminA}{arg\,min}
\DeclareMathOperator*{\argmaxA}{arg\,max}

\newcommand{\chapternote}[1]{{%
		\let\thempfn\relax
		\footnotetext[0]{\emph{#1}}
}}

\usepackage[margin=2.8cm]{geometry}

\newcounter{hypH}
\newenvironment{hypH}[1]
{
	{\vskip 2mm\noindent\refstepcounter{hypH}\textbf{(H\thehypH)}\quad #1}
	\vskip 2mm
}

\usepackage{setspace}
\begin{document}

\title{\bf {\Large Model-assisted estimation in high-dimensional settings for survey data}}

\author{Mehdi {\sc Dagdoug}$^{(1)}$,  Camelia {\sc Goga}$^{(1)}$ and  David  {\sc Haziza}$^{(2)}$ \\
	(1) Universit\'e de Bourgogne Franche-Comt\'e,\\ Laboratoire de Math\'ematiques de Besan\c con,  Besan\c con, FRANCE \\
	(2) University of Ottawa, Departement of Mathematics and Statistics,\\ Ottawa, CANADA
}

\date{\today}
\maketitle

\maketitle

\begin{abstract}
Model-assisted estimators have attracted a lot of attention in the last three decades. These estimators attempt to make an efficient use of auxiliary information available at the estimation stage. A working model linking the survey variable to the auxiliary variables is specified and fitted on the sample data to obtain  a set of predictions, which are then incorporated in the estimation procedures.  A nice feature of model-assisted procedures is that they maintain important design properties  such as  consistency and asymptotic unbiasedness irrespective of whether or not the working model is correctly specified. In this article, we examine several model-assisted estimators  from a design-based point of view and
in a high-dimensional setting, including penalized estimators and tree-based estimators. We conduct an extensive simulation study using data from the Irish Commission for Energy Regulation Smart Metering Project, in order to assess the performance of several model-assisted estimators in terms of bias and efficiency in this high-dimensional data set.
\end{abstract}

{\noindent  {\small {\em  Key words: Design consistency; Elastic net; Lasso; Random forest; Regression tree; Ridge regression.}
 } }

\section{Introduction}
Surveys conducted by national Statistical Offices (NSO) aim at estimating finite population parameters, which are those describing some aspect of the finite population under study. In this article, the interest lies in estimating the population total of a survey variable $Y$. Population totals can be estimated unbiasedly using the well-known Horvitz-Thompson estimator \citep{horvitz-thompson_1952}.  In the absence of nonsampling errors, the Horvitz-Thompson estimator is unbiased with respect to the customary design-based inferential approach, whereby the properties of estimators are evaluated with respect to the sampling design; e.g., see \cite{SarndalLivre}. However, Horvitz-Thompson type estimators may exhibit a large variance in some situations. The efficiency of the Horvitz-Thompson estimator can be improved by incorporating some auxiliary information,  capitalizing on the relationship between the survey variable $Y$ and a set of auxiliary variables $\mathbf{x}$.  The resulting estimation procedures, referred to as model-assisted estimation procedures, use a working model as a vehicle for constructing point estimators.  Model-assisted estimators remain design-consistent even if the working model is misspecified, which is a desirable feature. When the working model provides an adequate description of the relationship between $Y$ and $\mathbf{x}$, model-assisted estimators are expected to be more efficient than the Horvitz-Thompson estimator.

The class of model-assisted estimators include a wide variety of procedures, some of which have been extensively studied in the literature both theoretically and empirically. When the working model is the customary linear regression model, the resulting estimator is the well-known generalized regression estimator (GREG); e.g., \cite{sarndal_1980}, \cite{sarndal_wright_1984} and \cite{SarndalLivre}. Other works include model-assisted procedures based on generalized linear models \citep{lehtonen1998logistic, firth1998robust}, local polynomial regression \citep{breidt_opsomer_2000}, splines \citep{breidt_claeskens_opsomer_2005,goga_2005, mcconville_breidt_2013,goga_ruiz-gazen}, neural nets \citep{montanari_ranalli_2005}, generalized additive models \citep{opsomer2007model}, nonparametric additive models \citep{wang_wang_2011}, regression trees \citep{toth2011building, mcconville2019automated} and random forests \citep{dagdoug2020model}.

Due to the recent advances of information technology, NSOs have now access to a variety of data sources, some of which may exhibit a large number of observations on a large number of variables. So far, the properties of model-assisted estimator have been established under the customary asymptotic framework in finite population sampling \citep{isaky_fuller_1982} for which both the population size $N$ and the sample size $n$ increase to infinity, while assuming that the number of auxiliary variables $p$ is fixed. In other words, existing results require $n$ to be large relative to $p$. This framework is generally not adequate in the context of high-dimensional data sets as $p$ may be of the same order as $n$, or even larger, i.e., $p>n$. Therefore, a more appropriate asymptotic framework would let $p$ increase to infinity in addition to $N$ and $n$. \cite{cardot_goga_shehzad_2017} studied dimension reduction through principal component analysis and established the design consistency of the resulting calibration estimator. More recently, \cite{Ta2020} investigated the properties of the GREG estimator  from a model point of view and when $p$ is allowed to diverge.

The aim of this paper is to investigate the design properties of a number of model-assisted estimation procedures when $p$ is allowed to grow to infinity. We start by showing that, under mild assumptions on the sampling design and the auxiliary information, the GREG estimator is deign-consistent provided that $p^3/n$ goes to zero. 
In a high-dimensional setting,  one can recourse to penalization methods such as ridge, lasso and elastic net. We investigate the theoretical properties of model-assisted estimators based on these penalized methods and show that, under very mild regularity conditions,  these estimators are design-consistent provided that $p^3/n$ goes to zero. As we argue in Section 3, this rate can be improved if one is willing to make additional assumptions about the rate of convergence of the estimated regression coefficient. In particular, we lay out a set of additional conditions under which the model-assisted ridge estimator is consistent if $p/n$ goes to zero and moreover, is $\sqrt{n}$-consistent if $p=\mathcal O(n^a)$ with $a\in (0,1/2).$ Also, provided that the predictors orthogonal, we show that both the model-assisted lasso and elastic net estimators are consistent provided that $p/n$ goes to zero.

When the relationship between the study and the auxiliary information does not appear to be linear in the auxiliary variables, the problem of model-assisted estimation becomes more challenging, especially in a high-dimensional setting, a phenomenon known as the curse of dimensionality. Tree based methods that include regression trees and random forests tend to be efficient in these conditions. We establish the $L^1$-design consistency of some tree-based methods in a high-dimensional setting
without any condition on the number $p$ of auxiliary variables. Hence, these methods remain consistent even when $p$ is much larger than $n$, which is an attractive feature. To assess the performance of several model-assisted estimators in a high-dimensional setting, we conduct a simulation study using data from the Irish Commission for Energy Regulation Smart Metering Project. The data set consists of electricity consumption recorded every half an hour for a two-year period and for more than $6000$ households and businesses, leading to highly correlated data. Due to the high-dimensional feature, linear relationships are difficult to detect and even if we succeed, estimators based on linear models tend to breakdown. In this case, penalized and tree-based model-assisted estimators may provide good alternatives. However, an empirical comparison of  these model-assisted estimators in terms of bias and efficiency in a high-dimensional setting is currently lacking. We aim to fill this gap in the article.

The paper is organized as follows. In Section \ref{sec2}, we introduce the theoretical setup. In Section \ref{sec3}, we investigate the asymptotic design properties of several model-assisted estimators: the GREG estimator as well as estimators based on ridge regression, lasso and elastic net. Tree-based methods in a high-dimensional setting are considered in Section \ref{sec4}. Section \ref{sec5} contains an empirical comparison  to assess the performance of several model-assisted estimators in terms of bias and efficiency. We make some final remarks in Section \ref{sec6}. The technical details, including the proofs of some results, are relegated to the Appendix.

\section{The setup}\label{sec2}
Consider a finite population $U = \{1, 2, ..., N\}$ of size $N$. We are interested in estimating $t_y = \sum_{i \in U} y_i$, the population total of the survey variable $Y$. We select a sample  $S$ from $U$ according to a sampling design $\mathcal{P}(S)$ with first-order and second-order inclusion probabilities $\{\pi_i\}_{i \in U}$ and $\{\pi_{i\ell}\}_{i,\ell \in U}$, respectively. In the absence  of nonsampling errors, the Horvitz-Thompson estimator
\begin{equation} \label{ht}
\widehat{t}_{\pi} = \sum_{i \in S} \dfrac{y_i}{\pi_i}
\end{equation}
is design-unbiased for $t_y$ provided that $\pi_i >0$ for all $i \in U$; that is, $\mathbb{E}_{\rm p}(\widehat{t}_{\pi})= t_y,$ where $\mathbb{E}_p(\cdot)$ denotes the expectation operator with respect to the sampling design $\mathcal {P}(S)$.
In the sequel, unless stated otherwise, the properties of estimators are evaluated with respect to the design-based approach.
Under mild conditions  \citep{robinson_sarndal_1983, breidt_opsomer_2000}, it can be shown that the Horvitz-Thompson estimator $\widehat{t}_{\pi}$ is design-consistent for $t_y$.

At the estimation stage, we assume that a collection of auxiliary variables, $X_1, X_2, \ldots, X_p$, is recorded for all $i \in S$. Moreover, we assume that the corresponding population totals are available from an external source (e.g., a census or an administrative file).  Let $\bx_i = \left[\bx_{i1}, \bx_{i2}, \ldots, \bx_{ip}\right]^\top$ be the $\mathbf{x}$-vector associated with unit $i$. Also,  we denote by $\boldsymbol{X}_U = (\bx^{\top}_i)_{i\in U}$ the $N\times p$ design matrix and $\boldsymbol{X}_S=(\bx^{\top}_i)_{i\in S}$ its sample counterpart. 

Model-assisted estimation starts with postulating the following working model:
\begin{equation} \label{model}
	\xi : \ \ y_i = f(\bx_i) + \epsilon_i, \quad i\in U,
\end{equation}
 where $f(\bx_i) = \mathbb{E} \left[y_i \rvert \bx_i \right]$ and the errors $\epsilon_i$ are independent random variables such that $\mathbb{E}_\xi \left[\epsilon_i \rvert \bx_i\right] = 0$ and  $\V_\xi \left(\epsilon_i\rvert \bx_i\right) = \sigma^2$ for all $i \in U$. Although we assume an homoscedastic variance structure, our results can be easily extended to the case of unequal variances of the form $\V_\xi \left(\epsilon_i\rvert \bx_i\right) = \sigma^2\nu(\bx_i)$ for some known function $\nu(\cdot)$.

 The unknown function $f(\cdot)$ is estimated by $\widehat{f}(\cdot)$ from the sample data $(\bx_i, y_i)_{i \in S}$. 
 The fitted model is then used to construct the model-assisted estimator
 \begin{equation} \label{ma}
\widehat{t}_{ma} = \sum_{i \in U} \widehat{f}(\bx_i) + \sum_{i \in S} \dfrac{y_i - \widehat{f}(\bx_i)}{\pi_i},
 \end{equation}
 where $\widehat{f}(\bx)$ denotes the prediction at $\mathbf{x}$ under the working model (\ref{model}).
Whenever the predictor $\widehat{f}(\cdot)$ is sample dependent, the estimator $\widehat{t}_{ma}$ is design-biased, but can be shown to be asymptotically design-unbiased and design-consistent for a wide class of working models, as the population size $N$ and the sample size $n$ increase.


 \section{Least squares and penalized model-assisted estimators} \label{sec3}

 \subsection{The GREG estimator}
Suppose that the regression function $f(\cdot)$ is approximated by a linear combination of $X_j, j=1, \ldots, p$. The working model (\ref{model}) reduces to
\begin{equation}\label{lm}
\xi : \ \ y_i = \bx_i^\top \boldsymbol{\beta} + \epsilon_i, \quad i\in U,
\end{equation}
where $\boldsymbol{\beta} = \left[\beta_1, \ldots, \beta_p\right]^\top \in \mathbb{R}^p$ is a vector of unknown coefficients.  Under a hypothetical census, where we observe $y_i$ and $\mathbf{x}_i$ for all $i \in U,$ the vector $\boldsymbol{\beta}$  would be estimated by $\widetilde{\boldsymbol{\beta}}$ through the ordinary least square criterion at the population level:
 \begin{equation} \label{leastSquares}
\widetilde{\boldsymbol{\beta}} = \argminA_{\boldsymbol{\beta} \in \mathbb{R}^p}  \rvert\rvert\boldsymbol{y}_U  -  \boldsymbol{X}_U\boldsymbol{\beta} \rvert\rvert_2^2=\argminA_{\boldsymbol{\beta} \in \mathbb{R}^p} \sum_{i\in U}(y_i-\bx_i^\top \boldsymbol{\beta})^2,
 \end{equation}
where $\boldsymbol{y}_U=(y_i)_{i\in U}.$
Provided that the matrix $\boldsymbol{X}_U$ is of full rank, the solution to (\ref{leastSquares}) is unique and given by
 \begin{equation} \label{popBeta}
\widetilde{\boldsymbol{\beta}}  = \left(\boldsymbol{X}_U^\top \boldsymbol{X}_U\right)^{-1} \boldsymbol{X}^\top_U \boldsymbol{y}_U=\left(\sum_{i\in U}\bx_i\bx_i^\top\right)^{-1}\sum_{i\in U}\bx_iy_i.
 \end{equation}
In practice, the vector $\widetilde{\boldsymbol{\beta}} $ in (\ref{popBeta}) cannot be computed as the $y$-values are recorded for the sample units only.  An estimator of $\widetilde{\boldsymbol{\beta}},$ denoted by $\widehat{\boldsymbol{\beta}},$ is obtained from (\ref{popBeta}) by estimating each total separately using the corresponding Horvitz-Thompson estimator.   Alternatively, the estimator $\widehat{\boldsymbol{\beta}}$  can be obtained using the following weighted least square criterion at the sample level:
  \begin{equation} \label{sampleBetaEst}
\widehat{\boldsymbol{\beta}} = \argminA_{\boldsymbol{\beta} \in \mathbb{R}^p} \left(\boldsymbol{y}_S - \boldsymbol{X}_S\boldsymbol{\beta} \right)^\top \boldsymbol{\Pi}^{-1}_S \left(\boldsymbol{y}_S -  \boldsymbol{X}_S\boldsymbol{\beta} \right)^\top=\argminA_{\boldsymbol{\beta} \in \mathbb{R}^p} \sum_{i\in S}\frac{(y_i-\bx_i^\top \boldsymbol{\beta})^2}{\pi_i},
 \end{equation}
where $\boldsymbol{\Pi}_S = \text{diag}\left(\pi_i\right)_{i \in S}$ and $\boldsymbol{y}_S=(y_i)_{i\in S}$. Again, the solution to (\ref{sampleBetaEst}) is unique provided that $\boldsymbol{X}_S$ is of full rank and it is  given by
 \begin{equation}\label{sampleBeta}
\widehat{\boldsymbol{\beta}} = \left(\boldsymbol{X}_S^\top\boldsymbol{\Pi}^{-1}_S \boldsymbol{X}_S\right)^{-1} \boldsymbol{X}^\top_S \boldsymbol{\Pi}^{-1}_S \boldsymbol{y}_S = \left( \sum_{i \in S} \dfrac{ \bx_i \bx_i^\top}{\pi_i}\right)^{-1}  \sum_{i \in S} \dfrac{\bx_i y_i}{\pi_i}.
 \end{equation}
The prediction of $f(\cdot)$ at $\mathbf{x}$ under the working model (\ref{lm})  is $\widehat{f}_{\rm{lr}}(\bx) = \bx^\top \widehat{\boldsymbol{\beta}}$. Plugging $\widehat{f}_{\rm{lr}}(\cdot)$ in (\ref{ma}) leads to the well-known GREG estimator \citep{SarndalLivre}:
 \begin{eqnarray}\label{greg}
 	\widehat{t}_{\rm{greg}} &= & \sum_{i \in U} \widehat{f}_{\rm{lr}}(\bx_i) + \sum_{i \in S} \dfrac{y_i - \widehat{f}_{\rm{lr}}(\bx_i)}{\pi_i}\nonumber\\
	& =& \left(\sum_{i \in U} \bx_i\right)^{\top} \widehat{\boldsymbol{\beta}}+ \sum_{i \in S} \dfrac{y_i - \bx^{\top}_i\widehat{\boldsymbol{\beta}}}{\pi_i}.
\end{eqnarray}
  If the intercept is included in the working model,  the GREG estimator reduces to the population total of the fitted values $\widehat{f}_{lr}(\bx_i)=\bx_i^{\top} \widehat{\boldsymbol{\beta}};$ that is, $\widehat{t}_{\rm{greg}}=(\sum_{i\in U}\bx_i)^{\top} \widehat{\boldsymbol{\beta}}.$
Also, the GREG estimator can be written as a weighted sum of the sample $y$-values:
\begin{eqnarray}	
	\widehat{t}_{\rm{greg}}=\sum_{i \in S} w_{iS} y_i,
 \end{eqnarray}
 where
 $$w_{iS} =\displaystyle \frac{1}{\pi_i} \left\{ 1 + \left( \sum_{i \in U} \bx_i -\sum_{i \in S} \frac{\bx_i}{\pi_i} \right)^\top \left( \sum_{i \in S} \dfrac{ \bx_i \bx_i^\top}{\pi_i}\right)^{-1} \bx_i\right\}, \quad i\in S.$$
 These weights can be also obtained as the solution of a calibration problem \citep{deville_sarndal_1992}.  More specifically, the weights $w_{iS}$ are such that the generalized chi-square distance $\sum_{i\in S}(\pi_i^{-1}-w_{iS})^2/\pi_i^{-1}$ is minimized subject to the calibration constraints $\sum_{i \in S} w_{iS} \bx_i = \sum_{i\in U}\bx_i.$ This attractive feature may not be shared by model-assisted estimators derived under more general working models.

\subsection{Penalized least square estimators}

While model-assisted estimators based on linear regression working models are easy to implement, they tend to breakdown when the number of auxiliary variables $p$ is growing large. Also, when some of the predictors are highly related to each other, a problem known as multicolinearity, the ordinary least square estimator $\widetilde{\boldsymbol{\beta}}$ given by (\ref{popBeta}) may be highly unstable. As noted by \cite{hoerl_kennard_2000}, ``the worse the conditioning of $\boldsymbol{X}_U^{\top}\boldsymbol{X}_U,$ the
more $\widetilde{\boldsymbol{\beta}}$ can be expected to be too long and the distance from $\widetilde{\boldsymbol{\beta}}$ to $\boldsymbol{\beta}$ will tend to be large''. In  survey sampling, the effect of multicolinearity on the stability of point estimators has first been studied by  \cite{bardsley_chambers} under the model-based approach.  \cite{Chambers96} and \cite{rao_singh_1997} examined this problem in the context of calibration. These authors noted that, using a large number of calibration constraints, may lead to highly dispersed calibration weights, potentially resulting in unstable estimators.

In a classical \textit{iid} linear regression setting, penalization procedures such as ridge, lasso or elastic-net can be used to help circumvent some of the difficulties associated
with the usual least squares estimator $\widetilde{\boldsymbol{\beta}}$. Let $\widetilde{\boldsymbol{\beta}}_{\rm{pen}}$ be an estimator of $\boldsymbol{\beta}$ obtained through the penalized least square criterion at the population level:
 \begin{align} \label{GeneralOptSS2}
 \widetilde{\boldsymbol{\beta}}_{\rm{pen}} = \argminA_{\boldsymbol{\beta} \in \mathbb{R}^p} \sum_{i \in U}\left(y_i- \bx_i^{\top}\boldsymbol{\beta}\right)^2 +\sum_{\ell = 1}^t  \lambda_{\ell}\rvert \rvert \boldsymbol{\beta} \rvert \rvert^{\gamma_{\ell}}_{\nu_{\ell}},
 \end{align}
where $\lambda_{\ell}$, $\nu_{\ell}$  and $\gamma_{\ell}$ are positive real numbers, $||\cdot||_{_{\nu}}$ is a given norm and $t$ is a fixed positive integer representing the number of different norm constraints. The values of $\nu_{\ell},$ $\gamma_{\ell}$ and $t$ are typically predetermined.  The tuning parameters $\lambda_{\ell}$ control the strength of the penalty that one wants to impose on the norm of $\boldsymbol{\beta}$ and, most often,  $\lambda_{\ell}$ is chosen by cross-validation.  
The coefficients $\gamma_{\ell}$ and $\nu_{\ell}$ are specific to the penalization method. Hence, they have an impact on the properties of the resulting estimator $ \widetilde{\boldsymbol{\beta}}_{\rm{pen}}. $ Three special cases are considered below.

When $t = 1,$ $\gamma_1= 2$ and $\nu_1 = 2$, $\lambda_1=\lambda,$ the estimator is known as the ridge regression estimator \citep{hoerl_kennard_1970}:
   \begin{align*}
 \widetilde{\boldsymbol{\beta}}_{\rm{ridge}} = \argminA_{\boldsymbol{\beta} \in \mathbb{R}^p} \sum_{i \in U}\left(y_i- \bx_i^{\top}\boldsymbol{\beta}\right)^2 +\lambda\rvert \rvert \boldsymbol{\beta} \rvert \rvert^2_2,
 \end{align*}
where $\rvert \rvert \boldsymbol{\beta} \rvert \rvert^2_2=\sum_{j=1}^p\beta^2_j$ is the usual Euclidian norm of $\boldsymbol{\beta}.$ The solution is given explicitly by
 \begin{eqnarray}\label{ridge}
 \widetilde{\boldsymbol{\beta}}_{\rm{ridge}} &= & \left(\boldsymbol{X}_U^\top \boldsymbol{X}_U+\lambda \boldsymbol{I}_{p}\right)^{-1} \boldsymbol{X}^\top_U \boldsymbol{y}_U= \left( \sum_{i \in U} \bx_i \bx_i^\top + \lambda \boldsymbol{I}_{p}\right)^{-1}  \sum_{i\in U}\bx_i y_i,
 \end{eqnarray}
 where $\boldsymbol{I}_p$ denotes the $p\times p$ identity matrix.

 When $\gamma_1= 1,$ $\nu_1=1$ and $\lambda_1=\lambda,$ the estimator $ \widetilde{\boldsymbol{\beta}}_{\rm{pen}}$ is known as the lasso estimator \citep{tibshirani1996regression}:
   \begin{align} \label{lasso}
 \widetilde{\boldsymbol{\beta}}_{\rm{lasso}} = \argminA_{\boldsymbol{\beta} \in \mathbb{R}^p} \sum_{i \in U}\left(y_i- \bx_i^{\top}\boldsymbol{\beta}\right)^2 +\lambda\rvert \rvert \boldsymbol{\beta} \rvert \rvert_1,
 \end{align}
 where $\rvert \rvert \boldsymbol{\beta} \rvert \rvert_1=\sum_{j=1}^p|\beta_j|$ is the $L^1$-norm of $\boldsymbol{\beta}.$ As for the ridge, the lasso has
 the effect of shrinking the coefficients but, unlike the ridge, it
 can set some coefficients $\beta_j$ to zero.
Except when the auxiliary variables are orthogonal,  there is no closed-form formula for the lasso estimator $\widetilde{\boldsymbol{\beta}}_{\rm{lasso}}.$ 
In survey sampling, \cite{mcconville2017model} investigated the design-based properties of the lasso model-assisted estimator for fixed $p$.

The elastic-net estimator, that was suggested by \cite{zou2005regularization},    combines two norms: the euclidean norm $||\cdot||_2$ and the $L^1$ norm, $||\cdot||_1$. If, in (\ref{GeneralOptSS2}), we set $t=2,$ $\gamma_1 = 1$, $\nu_1 =1$,  $\gamma_2 = 2,$ $\nu_2 =2,$ $\lambda_1=\lambda\alpha$ and $\lambda_1=\lambda(1-\alpha)$, the resulting estimator is the elastic-net estimator, which can be viewed as a trade-off  between the ridge estimator and the lasso estimator, realizing variable selection and regularization simultaneously:
\begin{eqnarray*}
\widetilde{\boldsymbol{\beta}}_{\rm{en}}=\argminA_{\boldsymbol{\beta} \in \mathbb{R}^p} \sum_{i \in U}\left(y_i- \bx_i^{\top}\boldsymbol{\beta}\right)^2 +\lambda \left[\alpha \rvert \rvert \boldsymbol{\beta} \rvert \rvert_{1}+(1-\alpha) \rvert \rvert \boldsymbol{\beta} \rvert \rvert^{2}_2\right],
\end{eqnarray*}
for $\lambda>0$ and $\alpha\in [0,1]$ a parameter that is usually chosen using a grid of multiple values of $\alpha$.
The penalized  regression estimator $\widetilde{\boldsymbol{\beta}}_{\rm{pen}}$  in (\ref{GeneralOptSS2}) is unknown as the $y$-values are not observed for the non-sample units. To overcome this issue, we use the following weighted penalized least square criterion at the sample level:
 \begin{align}
 \widehat{\boldsymbol{\beta}}_{\rm{pen}} = \argminA_{\boldsymbol{\beta} \in \mathbb{R}^p} \sum_{i \in S}\frac{1}{\pi_i}\left(y_i- \bx_i^{\top} \boldsymbol{\beta}\right)^2 +\sum_{\ell = 1}^t  \lambda_{\ell}\rvert \rvert \boldsymbol{\beta} \rvert \rvert^{\gamma_{\ell}}_{\nu_{\ell}}.
 \end{align}
%
%
A model-assisted estimator based on a penalized regression procedure is obtained from (\ref{ma}) by replacing $\widehat{f}(\bx)$ with $\widehat{f}_{\rm{pen}}(\bx)=\bx^\top \widehat{\boldsymbol{\beta}}_{\rm{pen}},$ leading to
 \begin{eqnarray}\label{est_pen}
 \widehat{t}_{\rm pen} &= & \sum_{i \in U} \widehat{f}_{\rm{pen}}(\bx_i) + \sum_{i \in S} \dfrac{y_i - \widehat{f}_{\rm{pen}}(\bx_i)}{\pi_i}\nonumber\\
 &=& \left(\sum_{i \in U} \bx^{\top}_i\right) \widehat{\boldsymbol{\beta}}_{\rm{pen}}+ \sum_{i \in S} \dfrac{y_i - \bx^{\top}_i\widehat{\boldsymbol{\beta}}_{\rm{pen}}}{\pi_i},
\end{eqnarray}
where $\widehat{\boldsymbol{\beta}}_{\rm{pen}}$ is a generic notation used to denote the estimated regression coefficient obtained through either lasso, ridge or elastic net.
Unlike the GREG estimator,
$ \widehat{t}_{\rm greg},$ the penalized model-assisted estimator is sensitive to unit change of the $X$-variables because $ \widehat{\boldsymbol{\beta}}_{\rm{pen}}$ is sensitive to unit change. This is why, as in the classical regression setting, standardization of the $X$-variables is recommended before computing $ \widehat{\boldsymbol{\beta}}_{\rm{pen}}. $ If the intercept is included in the model, then it is usually left un-penalized.

\begin{remark}
In the case of ridge regression, the estimator $\widehat{\boldsymbol{\beta}}_{\rm{ridge}}$ is given by
 \begin{eqnarray}\label{ridge_beta}
 \widehat{\boldsymbol{\beta}}_{\rm{ridge}} =  \left(\boldsymbol{X}_S^\top\boldsymbol{\Pi}^{-1}_S \boldsymbol{X}_S+\lambda \boldsymbol{I}_{p}\right)^{-1} \boldsymbol{X}^\top_S\boldsymbol{\Pi}^{-1}_S \boldsymbol{y}_S= \left( \sum_{i \in S}\frac{\bx_i \bx_i^\top}{\pi_i}  + \lambda \boldsymbol{I}_{p}\right)^{-1}  \sum_{i\in S}\frac{\bx_i y_i}{\pi_i}.
 \end{eqnarray}
 Using (\ref{ridge_beta}) in (\ref{est_pen}) leads to the ridge model-assisted estimator $\hat t_{\rm{ridge}}$ that can be expressed as a weighted sum of sampled $y$-values, $\hat t_{\rm{ridge}}=\sum_{i\in S}w_{iS}(\lambda)y_i$ with weights given by
 $$w_{iS}(\lambda) =\frac{1}{\pi_i}\left\{ 1 + \left( \sum_{i \in U} \bx_i -\sum_{i \in S} \frac{\bx_i}{\pi_i} \right)^\top \left( \sum_{i \in S} \dfrac{ \bx_i \bx_i^\top}{\pi_i}+\lambda \boldsymbol I_p\right)^{-1} \bx_i\right\}, \quad i\in S.$$
 These weights can also be obtained through a penalized calibration problem. It can be shown  that they minimize  the penalized generalized chi-square distance, $\sum_{i\in S}(\pi_i^{-1}-w_{iS})^2/\pi_i^{-1}+\lambda^{-1}||\sum_{i \in U} \bx_i -\sum_{i \in S} \bx_i/\pi_i||_2^2$ \citep{Chambers96, beaumont_bocci_2008}. If some $X$-variables are left un-penalized in (\ref{GeneralOptSS2}), the resulting weights ensure consistency between the survey estimates and their corresponding population totals associated with these variables.
 \end{remark}

\subsection{Consistency of the GREG and penalized GREG estimators in a high-dimensional setting}

We adopt the asymptotic framework of \cite{isaky_fuller_1982} and consider an increasing sequence of embedded finite populations $\{U_{v}\}_{v \in \mathbb{N}}$ of size $\{N_v\}_{v \in \mathbb{N}}$. In each finite population $U_v$, a sample,  of size $n_v,$ is selected according to a sampling design $\mathcal{P}_v(S_v)$ with first-order inclusion probabilities $\pi_{i, v}$ and second-order inclusion probabilities $\pi_{i\ell, v}$. While the finite populations are considered to be embedded, we do not require this property to hold for the samples $\{S_v\}_{v \in\mathbb N}$. This asymptotic framework assumes that $v$ goes to infinity, so that both the finite population sizes $\{N_v\}_{v \in\mathbb N}$, the samples sizes $\{n_v\}_{v \in\mathbb N}$ and the number of auxiliary variables $\{p_v\}_{v \in\mathbb N}$ go to infinity. To improve readability, we shall use the subscript $v$ only in the quantities $U_v, N_v$, $n_v$ and $p_v$; for instance, quantities such as $\pi_{i,v}$ shall be simply denoted by $\pi_i$.

The following assumptions are required to establish the consistency of the  GREG and penalized GREG estimators in a high-dimensional setting.

\hypH{We assume that there exists a positive constant $C_1$ such that  $N_v^{-1} \sum\limits_{i \in U_v} y_i^2 < C_1.  $  \label{H1}}

\hypH{We assume  that $ \lim\limits_{v \to \infty}\dfrac{n_v}{N_v} = \pi \in (0;1)$.\label{H2}}

\hypH{ There exist a positive constant $c$ such that $\min\limits_{i \in U_v} \pi_i \geqslant c> 0;$ also, we assume that $\limsup\limits_{v \to \infty} n_v \max\limits_{i \neq \ell \in U_v} | \pi_{i\ell} - \pi_i \pi_{\ell}| < \infty$.
	\label{H3}}
\hypH{We assume that there exists a positive constant $C_2$ such that, for all $i \in U_v$, $\rvert\rvert \bx_i \rvert\rvert_2^2 \leq C_2p_v$, where $||\cdot||_2$ denotes the usual Euclidian norm. \label{Hnew}}


\hypH{
We assume that $ ||\widehat{\boldsymbol{\beta}}||_1=\mathcal{O}_{\rm p}(p_v),$ where $\widehat{\boldsymbol{\beta}}$ is the least square estimator given in (\ref{sampleBeta}) and $||\cdot||_1$ denotes the $L^1$ norm. \label{H4}	}


The assumptions (H\ref{H1}), (H\ref{H2}) and (H\ref{H3}) were used by \cite{breidt_opsomer_2000} in a nonparametric setting and similar assumptions were used by \cite{robinson_sarndal_1983} to establish the consistency of the GREG estimator in a fixed dimensional setting. These assumptions hold for many usual sampling designs such as simple random sampling without replacement, stratified designs  \citep{breidt_opsomer_2000}, or high-entropy sampling designs. Assumptions (H\ref{Hnew}) and (H\ref{H4})  can be viewed, respectively, as extensions of Assumption A.1 and Assumption A.3  in \cite{robinson_sarndal_1983} to $p_v$-dimensional vectors with $p_v$ growing to infinity. Assumption (H\ref{H4}) is not very restrictive in this high-dimensional setting, it requires that components of $\widehat{\boldsymbol{\beta}}$ are all bounded.  When $p_v$ is fixed, then our assumptions essentially reduce to those of \cite{robinson_sarndal_1983}. 

\begin{result} \label{res1}
	Assume (H\ref{H1})-(H\ref{H4}). Consider a sequence of GREG estimators $\{\widehat{t}_{greg}\}_{v\in \mathbb N}$ of $t_y$. Then, 
	\begin{align*}
	\dfrac{1}{
		N_v} ( \widehat{t}_{\rm{greg}} -t_y)  = \mathcal{O}_{\rm p}\bigg(\sqrt{\dfrac{p^3_v}{n_v}}\bigg). 
	\end{align*}
If the numbers of auxiliary variables $\{p_v\}_{v\in \mathbb N}$ and the sample sizes $\{n_v\}_{v\in \mathbb N}$ satisfy $p_v^3 / n_v =o(1)$, then $N_v^{-1}(\widehat{t}_{greg} -t_y)  = o_{\rm p}(1).$
\end{result}
 	The $\sqrt{n}$-consistency obtained by \cite{robinson_sarndal_1983} is a special case of Result \ref{res1} with  $p_v = \mathcal{O}(1)$. Result \ref{res1} highlights the fact that the rate of convergence decreases as the number of auxiliary variables $p_v$ increases. Yet, this result guarantees the existence of a consistent GREG estimator, even when the number of auxiliary variables is allowed to diverge.  An improved consistency rate for $\widehat{t}_{\rm{greg}}$
 	 may be obtained if the usual euclidean norm
 	   is used instead of $L^1$-norm.
 	   Establishing the rate of convergence of the sampling error $\widehat{\boldsymbol{\beta}}-\widetilde{\boldsymbol{\beta}}$ may also be utilized to obtain a lower consistency rate for $\widehat{t}_{\rm{greg}}$.  This is beyond of the scope of this article.
 	
The next result establishes the design-consistency of model-assisted penalized regression estimators. The proof is similar to that of result \ref{res1} and is given in the Appendix. 
	


 \begin{result} \label{res2}
 Assume (H\ref{H1})-(H\ref{H4}). Consider a sequence of penalized model-assisted estimators $\{\widehat{t}_{\rm pen}\}_{v\in \mathbb N}$ of $t_y$ obtained by either ridge, lasso or elastic-net. Then,
 \begin{align*}
\dfrac{1}{
		N_v} ( \widehat{t}_{\rm{pen}} -t_y)  = \mathcal{O}_{\rm p}\bigg(\sqrt{\dfrac{p^3_v}{n_v}}\bigg). 
		\end{align*}
If the numbers of auxiliary variables $\{p_v\}_{v\in \mathbb N}$ and the sample sizes $\{n_v\}_{v\in \mathbb N}$ satisfy $p_v^3 / n_v =o(1)$, then $N_v^{-1}(\widehat{t}_{pen} -t_y)  = o_{\rm p}(1).$
\end{result}
The above result makes no use of the asymptotic convergence rate of $\widehat{\boldsymbol{\beta}}_{\rm{pen}}$ which depends on the penalization method. For example, if one can establish that $\rvert \rvert \widehat{\boldsymbol{\beta}}_{\rm{pen}}  \rvert \rvert _1=  \mathcal{O}_{\rm p} (\gamma_v)$, then $N^{-1}_v (\widehat{t}_{\rm pen} -t_y)= \mathcal{O}_{\rm p} (\gamma_v\sqrt{p_v/n_v}). $  Alternatively, improved consistency rates of $\widehat{t}_{\rm{pen}}$ may be obtained if one can establish the extent of the sampling error $\widehat{\boldsymbol{\beta}}_{\rm{pen}}-\widetilde{\boldsymbol{\beta}}_{\rm{pen}}$ in a high-dimension setting. However, obtaining these improved rates requires additional assumptions, while Result \ref{res2} is obtained under relatively mild assumptions.

Next, we show that, under additional assumptions on the auxiliary variables, the model-assisted ridge estimator is $L^1$ design-consistent for $t_y$ if $p_v/n$ goes to zero and that it has the usual $\sqrt{n}$-consistency rate if $p_v=O(n_v^{a})$ with $0<a<1/2,$ which constitutes a significant improvement over Result \ref{res2}.

 \begin{result}\label{result_consist_ridge}
 Assume (H\ref{H1})-(H\ref{H3}). Also, assume that there exists a positive constant $\tilde C$ such that  $\lambda_{max}(\boldsymbol{X}_{U_v}^{\top}\boldsymbol{X}_{U_v})\leqslant \tilde C N_v, $ where $\lambda_{max}(\boldsymbol{X}_{U_v}^{\top}\boldsymbol{X}_{U_v})$ is the largest eigenvalue of $\boldsymbol{X}_{U_v}^{\top}\boldsymbol{X}_{U_v}. $ Assume also that $N_v/\lambda_v=O(1).$
 \begin{enumerate}
 \item Then, there exists a positive constant $C$ such that $
 \mathbb{E}_p \left[\rvert \rvert \widehat{\boldsymbol{\beta}}_{\rm{ridge}}  \rvert \rvert _2^2 \right] \leqslant C$
and
 \begin{align*}
	\dfrac{1}{
		N_v} \mathbb{E}_p \bigg\rvert \widehat{t}_{\rm{ridge}} -t_y\bigg\rvert  = \mathcal{O} \bigg(\sqrt{\frac{p_v}{n_v}}\bigg).
	\end{align*}
	If the numbers of auxiliary variables $\{p_v\}_{v\in \mathbb N}$ and the sample sizes $\{n_v\}_{v\in \mathbb N}$ satisfy $p_v / n_v =o(1)$, then $N_v^{-1}\mathbb{E}_p \rvert \widehat{t}_{\rm ridge} -t_y\rvert =o(1).$ 
\item $\mathbb{E}_{\rm{p}}(\rvert \rvert\widehat{\boldsymbol{\beta}}_{\rm{ridge}} -\tilde{\boldsymbol{\beta}}_{\rm{ridge}}\rvert \rvert_2^2)=\mathcal{O}(p_v/n_v). $ If  $p_v / n_v =o(1),$ then $\mathbb{E}_{\rm{p}}(\rvert \rvert\widehat{\boldsymbol{\beta}}_{\rm{ridge}} -\tilde{\boldsymbol{\beta}}_{\rm{ridge}}\rvert \rvert_2^2)=o(1). $
\item We have the following asymptotic equivalence:
$$
\frac{1}{N_v}\left(\widehat{t}_{\rm{ridge}} -t_y\right)=\frac{1}{N_v}\left(\widehat{t}_{\rm{diff,\lambda}} -t_y\right)+\mathcal{O}_{\rm p}\left(\frac{p_v}{n_v}\right)
$$
where $$\widehat{t}_{\rm{diff,\lambda}}=\sum_{i\in S_v}y_i/\pi_i-\left(\sum_{i\in S_v}\bx_i/\pi_i-\sum_{i\in U_v}\bx_i\right)^{\top}\tilde{\boldsymbol{\beta}}_{\rm{ridge}} $$
and
\begin{align*}
	\dfrac{1}{
		N_v} \mathbb{E}_p \bigg\rvert \widehat{t}_{\rm{ridge}} -t_y\bigg\rvert  =\mathcal{O} \bigg(\frac{1}{\sqrt{n_v}}\bigg)+ \mathcal{O} \bigg(\frac{p_v}{n_v}\bigg).
	\end{align*}
If $p_v=\mathcal O(n^a_v)$ with $0<a<1/2,$ then
$$
\frac{1}{N_v}\mathbb{E}_p \bigg\rvert \widehat{t}_{\rm ridge} -t_y\bigg\rvert =\mathcal{O} \bigg(\frac{1}{\sqrt{n_v}}\bigg).
$$
\end{enumerate}
\end{result}
The case of  model-assisted estimators based on lasso and elastic-net is more intricate. This is due to the fact that both estimators involve the $L^1$-norm. As a result, a closed-form of these estimators cannot be obtained. However, if the predictors are orthogonal, a closed-form expression exists for the lasso and elastic-net estimators and improved consistency rates can be obtained; see Proposition \ref{lasso_orthogon} below. The case of non-orthogonal predictors is more challenging and is beyond the scope of this article.


\begin{proposition}\label{lasso_orthogon}
Suppose assumptions (H\ref{H1})-(H\ref{H3}) and that the sampling design and the $X$-variables are such that the columns of $\boldsymbol{\Pi}^{-1/2}_{S_v}\mathbf{X}_{S_v}$ are orthogonal. Suppose also  that there exist a positive quantities  $C_3, C_4$ such that $\mbox{max}_{j=1, \ldots, p_v}N^{-1}_v\sum_{i\in U_v}x^4_{ij}\leq C_3<\infty$ and
$\mbox{min}_{j=1, \ldots, p_v}N^{-1}_v\sum_{i\in U_v}x^2_{ij}\geq C_4>0.$ Then $
N^{-1}_v( \widehat{t}_{\rm{greg}} -t_y) = \mathcal{O}_{\rm p}(\sqrt{p_v/n_v})
$
and
$
N^{-1}_v ( \widehat{t}_{\rm{pen}} -t_y)  = \mathcal{O}_{\rm p}(\sqrt{p_v/n_v}),$ where $\widehat{t}_{\rm{pen}}$ denotes the lasso or the elastic-net estimator.
\end{proposition}


 \section{Tree-based methods} \label{sec4}

 The penalization procedures considered in Section \ref{sec3} required the $\mathbf{x}$-vector of auxiliary variables to be available for $i \in S$ and the corresponding vector of population totals $\sum_{i \in U}\mathbf{x}_i$. In other words, the individual values of $\mathbf{x}_i, \ i \in U-S,$ were not needed. Also, the GREG estimator and the model-assisted estimators based on ridge regression, lasso and elastic net, were all based on a linear working model. In practice, the relationship between the survey variable $Y$ and  $\mathbf{x}$  may be nonlinear and one may have access to complete auxiliary information, which arises when the $\mathbf{x}$-vector is recorded for all $i \in U$. In this case, we may consider additional model-assisted estimators based on nonparametric procedures; e.g., kernel or spline regression \citep{breidt_opsomer_2000, breidt_claeskens_opsomer_2005, goga_2005, mcconville_breidt_2013}. However, these methods are vulnerable to the so-called curse of dimensionality, which arises when $p$ is large.
 The reader is referred to \citep[chapter 1]{hastie_tibshirani_friedman_2011} and \citep[chapter 2]{gyorfi2006distribution} for a discussion.
  In contrast, tree-based methods, that include regression trees and random forests as special cases, may prove useful when the number of auxiliary variables, $p,$ is large.
 \cite{mcconville2019automated} and \cite{dagdoug2020model} studied model-assisted estimators based on regression trees and random forests under the customary framework that treats the dimension of the $\mathbf{x}$-vector, $p,$ as fixed.
%

 \subsection{Model-assisted estimators based on regression trees}

Regression trees is one of the most popular statistical learning methods which assumes that the regression function $f$ has the form
\begin{eqnarray}
f(\mathbf{x})= \sum_{j=1}^T\mathds{1} \left(\bx \in A_j\right)\beta_j,\label{f_tree}
\end{eqnarray}
where $A_1, A_2, \ldots, A_T,$ are non-overlapping regions of $\mathbb{R}^p$ built recursively from the sample data $\{(y_i,\mathbf{x}_i)\}_{i\in S}$ and by using some optimization criterion for better prediction of $Y.$  A regression tree can be viewed as a linear regression model based on the set of predictors $Z_j(\bx)=\mathds{1} \left(\bx \in A_j\right)$ for $j=1, \ldots, T.$ 
The regions $A_1, \ldots, A_T$  are highly dependent on the sample data in terms of shape and number. Given $A_1, A_2, \ldots, A_T,$  the unknown parameters $\beta_j$ from (\ref{f_tree}) are estimated by  $\widehat{\boldsymbol{\beta}}=(\widehat{\beta_j})_{j=1}^T$ determined using the following weighted least-square criterion at the sample level:
\begin{eqnarray}
\widehat{\boldsymbol{\beta}}&=&\argminA_{\boldsymbol{\beta} \in \mathbb{R}^T}\sum_{i\in S}\frac{1}{\pi_i}(y_i-\mathbf{z}^{\top}_i\widehat{\boldsymbol{\beta}})^2,\label{beta_tree}
\end{eqnarray}
where $\mathbf{z}_i=(Z_j(\bx_i))_{j=1}^T=(\mathds{1} \left(\bx_i \in A_j\right))_{j=1}^T.$  The criterion (\ref{beta_tree}) is different from  the one used in the customary iid setup as it incorporates the sampling weights $\pi_i^{-1}$.  The solution to (\ref{beta_tree}) is given by
\begin{eqnarray}
\widehat{\boldsymbol{\beta}}=\left(\boldsymbol{Z}^{\top}_S\boldsymbol{\Pi}_{S}^{-1}\boldsymbol{Z}_S\right)^{-1}\boldsymbol{Z}^{\top}_S\boldsymbol{\Pi}_S^{-1}\boldsymbol{y}_S=\left(\sum_{i\in S}\frac{\mathbf{z}_i\mathbf{z}^{\top}_i}{\pi_i}\right)^{-1}\sum_{i\in S}\frac{\mathbf{z}_iy_i}{\pi_i},
\label{beta_tree2}
\end{eqnarray}
where $\boldsymbol{Z}_S=(\mathbf{z}^{\top}_i)_{i\in S}$ and $\boldsymbol{\Pi}_{S}=\mbox{diag}(\pi_i)_{i\in S}$.  Each unit $i\in S$ belongs to one and only one region. As a result, the matrix   $\boldsymbol{Z}^{\top}_S\boldsymbol{\Pi}_S^{-1}\boldsymbol{Z}_S$ is diagonal  with  elements equal to $\sum_{i\in S}\frac{1}{\pi_i}\mathds{1} \left(\bx_i \in A_j\right)$ for $j=1, \ldots T.$ Usually, most algorithms leading to the regions $A_1, A_2, \ldots, A_T,$  use a stopping criterion. The most common consists of specifying a minimum number, $n_0,$ of observations in each region. To avoid unduly large nodes, the algorithm also ensures that the number of observations in each node does not exceed $2n_0-1.$
Therefore, $\sum_{i\in S}\frac{1}{\pi_i}\mathds{1} \left(\bx_i \in A_j\right)\geq n_0>0$ for all $j=1, \ldots, T$ and the matrix $\boldsymbol{Z}^{\top}_S\boldsymbol{\Pi}_S^{-1}\boldsymbol{Z}_S$ is non-singular for all samples $S.$ The vector $\widehat{\boldsymbol{\beta}}$ is always well-defined and, using (\ref{beta_tree2}), we obtain
$$
\widehat{\boldsymbol{\beta}}=(\widehat{\beta}_j)_{j=1}^{T}=\left(\frac{\sum_{i\in S}\frac{1}{\pi_i}\mathds{1} \left(\bx_i \in A_j\right)y_i}{\sum_{i\in S}\frac{1}{\pi_i}\mathds{1} \left(\bx_i \in A_j\right)}\right)_{j=1}^T.
$$
The estimator $\widehat\beta_j$ is simply the weighted mean of the $y$-values associated with sample units falling in region $A_j. $


The prediction  of $f$ at $\mathbf{x}$ is then given by
\begin{equation} \label{treee}
\widehat{f}_{\rm tree} (\bx) =\sum_{j=1}^{T} \mathds{1} \left(\bx \in A_j\right)\widehat{\beta}_j
\end{equation}
and a model-assisted estimator based on regression trees is given by
\begin{eqnarray}\label{cart}
\widehat{t}_{tree} &= & \sum_{i \in U} \widehat{f}_{tree}(\bx_i) + \sum_{i \in S} \dfrac{y_i - \widehat{f}_{tree}(\bx_i)}{\pi_i}\nonumber\\
 &=& \sum_{i \in U} \mathbf{z}^{\top}_i \widehat{\boldsymbol{\beta}} + \sum_{i \in S} \dfrac{y_i - \mathbf{z}^{\top}_i \widehat{\boldsymbol{\beta}}}{\pi_i},
\end{eqnarray}
where $\widehat{f}_{tree}(\cdot)$ is defined in (\ref{treee}). As noted by \cite{mcconville2019automated}, the form of the estimator (\ref{cart}) is similar to that of a post-stratified estimator with sample-based post-strata $A_j, \ j=1, \ldots, T$.  However, unlike the customary post-stratified estimator which requires the population counts only, the estimator  $\widehat{t}_{tree}$ requires complete auxiliary information.


It can be shown that (\ref{cart}) can be expressed as a projection-type estimator.  Let $\mathbf{1}_T$ be the $T$-dimensional vector of ones. We have $\mathbf{1}^{\top}_T\mathbf{z}_i=1$ for all $i\in U$ since the vector $\mathbf{z}_i=(\mathds{1} \left(\bx_i \in A_j\right))_{j=1}^T$ contains only one coordinates equal to one and the rest are zero as $\mathbf{x}_i$ belongs to only one region $A_j$. It follows that the second term on the right hand-side of (\ref{cart}) is equal to zero since
$$
\sum_{i\in S}\frac{\mathbf{z}^{\top}_i \widehat{\boldsymbol{\beta}}}{\pi_i}=\mathbf{1}^{\top}_T\left(\sum_{i\in S}\frac{\mathbf{z}_i\mathbf{z}^{\top}_i }{\pi_i}\right)\widehat{\boldsymbol{\beta}}=\mathbf{1}^{\top}_T\sum_{i\in S}\frac{\mathbf{z}_iy_i}{\pi_i}=\sum_{i\in S}\frac{y_i}{\pi_i}.
$$
As a result, the estimator $\widehat{t}_{tree}$ reduces to
\begin{eqnarray}
\widehat{t}_{tree}=\sum_{i \in U} \mathbf{z}^{\top}_i \widehat{\boldsymbol{\beta}}.
\end{eqnarray}




Several regression tree algorithms may be used to obtain the set $\mathcal{A}=\{A_1, \ldots, A_T\}$ of non-overlapping regions of  $\mathbb{R}^p$. One popular algorithm is the CART algorithm suggested by \cite{breiman1984classification}.
The CART algorithm recursively searches for the splitting variable and the splitting position (i.e., the coordinates on the predictor space where to split) leading to the greatest possible reduction of the residual mean of squares before and after the splitting. More specifically, let $A$ be a region or node and let $\#(A)$ be the number of units falling in $A. $ A split in $ A$ consists in finding a pair $(l,z),$ where $l$ is the variable coordinate, whose value lies between $1$ and $p$  and $z$ denotes the position of the split along the $l$th coordinate, within the limits of $A. $ Let $\mathcal{C}_{A}$ be set of all possible pairs $(l,z)$ in $A$.
The splitting process is performed by searching for the best split $(l^*,z^*)$ in the following sense:
\begin{eqnarray*} \label{opt1}
\left(l^*, z^*\right) &= &\argmaxA_{\left(l,z\right) \in \mathcal{C}_{A}} L(l,z)\quad \mbox{with}\nonumber\\
\hspace{-0.5cm} L(l,z) &=& \dfrac{1}{\#(A)} \sum_{i \in S} \mathds{1}(\mathbf{x}_i \in A) \left\{ \left(y_i - \bar{y}_A\right)^2 -  \left(y_i -\bar{y}_{A_{L}} \mathds{1}(X_{i\ell} < z) -\bar{y}_{A_{R}}\mathds{1}(X_{i\ell} \geqslant z) \right)^2 \right\},
\end{eqnarray*}
where $X_{ij}$ is the measure of $j$-th variable $X_j$ for the $i$th individual, $A_{L} = \left\{ \textbf{X} \in A ; \textbf{X}_{l} < z \right\}$, $A_{R} = \left\{ \textbf{X} \in A ; \textbf{X}_{l} \geqslant z \right\}$ with $\textbf{X}_{l}$ the $\ell$th coordinate of $X;$ $\bar{y}_A$  is the mean of $y_i$ for those units $i$ such that $\bx_i \in A$. The splitting algorithm stops when an additional split in a cell would cause the subsequent terminal nodes to have less than $n_0$ observations. Also, the algorithm ensures that the maximum number of units is less that $2n_0-1$ in each terminal node.  The tree model-assisted estimator suggested by \cite{mcconville2019automated} is based on a different regression tree algorithm, but both share many properties since the splitting criterion has a limited impact on the theoretical properties of the resulting model-assisted estimators.

To establish the consistency of $\widehat{t}_{tree}$ given by (\ref{cart}), we need the following additional assumption:

 \hypH{We assume  that $\mathbb{E}_{\rm p} \left[I_i I_{\ell} \rvert \mathcal A\right] = \pi_{i\ell} + r_{i\ell}$, with $\mathbb{E}_{\rm p}(\max\limits_{i, \ell \in U_v} |r_{i\ell}|)= \mathcal O (n_v^{-1})$.\label{H5}}

 Assumption (H\ref{H5}) \citep{toth2011building}   requires the sample data to be regular (i.e. without extreme observations). This means that the partition $\mathcal A$ has a decreasing influence on the probability that two elements belong to the sample. 
 The reader is referred to \cite{toth2011building}, \cite{mcconville2019automated} and \cite{dagdoug2020model} for more details about this assumption.

 We show below that the tree-based model-assisted is $L^1$ design-consistent by using the analogy between linear regression and regression trees and adapting the proof of Robinson and S\"arndal to a high-dimensional setting, which is different from the method used by \cite{mcconville2019automated}.
 \begin{result} \label{res3}
 	Assume (H\ref{H1})-(H\ref{H3}), (H\ref{H5}). Consider a sequence of tree model-assisted estimators $\{\widehat{t}_{tree}\}_{v\in \mathbb N}$ for $t_y$. We assume that the stopping criterion requires that there are at least  $n_{0v}$ units in each region $A_j, j=1, \ldots, T_v.$ Then, 
 	\begin{align*}
 	\mathbb{E}_p \bigg[ \dfrac{1}{
 		N_v} \bigg\rvert \widehat{t}_{\rm{tree}} -t_y\bigg\rvert  \bigg] = \mathcal{O} \bigg(\dfrac{1}{\sqrt{n_{0v}}}\bigg).
 	\end{align*}
 \end{result}
It is worth pointing out that Result \ref{res3} holds for a  broader class of algorithms, namely any regression tree where the splitting criterion requires a minimal number of elements in each terminal node. When there is no minimal number of element in each terminal node, the estimator may be unstable and additional assumptions would be needed. 


\subsection{Model-assisted estimators based on random forests}

While regression trees are simple to interpret, their predictive performances are often criticized. If the number of elements in each terminal node is too small, the predictions may suffer from a high-model variance.
 When the number of observations in each terminal node is large, the model variance decreases but the model bias increases.  Random forests \citep{breiman2001random}  have been suggested to improve on the efficiency of regression trees. Random forests are widely recognized for their predictive performances in a wide variety of scenarios. In a survey sampling framework, \cite{dagdoug2020model} studied the properties of model-assisted estimators based on random forests.

Random forests is an ensemble method based on $B$ randomized regression trees. A predictor $\widehat{f}$ is said to be randomized if there exists a random variable $\theta$ such that $\widehat{f}(\bx) = \widehat{f}(\bx, \theta)$. In other words, a randomized predictor is a predictor which may use a random variable to yield its prediction; the reader is referred to \cite{dagdoug2020model} and \cite{biau2008consistency} for more details. There exist several random forests algorithms which either differ in the way the regression trees are built or in the type of randomization they use. We now describe the random forest algorithm suggested by \cite{breiman2001random}.

Let $\Theta_1, \Theta_2, ..., \Theta_B$ be $B$ random variables used to randomize the regression trees. The main steps of the algorithm  can be summarized as follows:
\begin{enumerate}[Step 1:]
	\item Consider \textit{B} bootstrap data sets $D_1(\Theta_1), D_2(\Theta_2), ...,  D_{B}(\Theta_B)$  by selecting  $n$ pairs $(y_i, \bx_i)$ from $ \left\{(y_i,\bx_i)\right\}_{i \in S}$ with replacement.
	\item On each bootstrap data set $D_b(\Theta_b)$ for $b=1, \ldots, B,$ fit a regression tree and determine the prediction $\widehat{f}_{tree}^{(b)}$ of the unknown regression function $f$ from model (\ref{model}). For each regression tree, only $p_0$ variables randomly chosen among the $p$ variables are considered to search for the best split in (\ref{opt1}). Typical values of $p_0$ include $p_0=p/3$ and $p_0=\sqrt{p}. $ A stopping criterion is also used for obtaining the terminal nodes of each of the $B$ trees. For example, the splitting algorithm stops when a minimum number $n_0$ of units is reached in each terminal node.
	\item  The predicted value at $\bx$ is obtained by averaging the predictions at the point $\bx$ from each of the $B$ regression trees:
	\begin{equation} \label{rf1}
	\widehat{f}_{rf}(\bx) = \dfrac{1}{B} \sum_{b = 1}^B \widehat{f}_{tree}^{(b)}(\bx, \Theta_b).
	\end{equation}
\end{enumerate}
The random forest model-assisted estimator of $t_y$ is obtained by plugging $\widehat{f}_{rf}$ in (\ref{ma}), which leads to
 \begin{equation}\label{rf}
	\widehat{t}_{rf} =  \sum_{i \in U} \widehat{f}_{rf}(\bx_i) + \sum_{i \in S} \dfrac{y_i - \widehat{f}_{rf}(\bx_i)}{\pi_i},
	\end{equation}
where $\widehat{f}_{rf}(\bx_i)$ is given by (\ref{rf1}).

\begin{result} \label{res4}
	Assume (H\ref{H1})-(H\ref{H3}). Assume also that assumption (H\ref{H5}) holds for all partitions $\mathcal A_b, b=1, \ldots, B$ created for each of  the $B$ randomized trees. Consider a sequence of random forest model-assisted estimators $\{\widehat{t}_{rf}\}_{v\in \mathbb N}$ for $t_y.$ Then,
	\begin{align*}
	\mathbb{E}_p \bigg[ \dfrac{1}{
		N_v} \bigg\rvert \widehat{t}_{rf} -t_y\bigg\rvert  \bigg] = \mathcal{O} \bigg(\dfrac{1}{\sqrt{n_{0v}}}\bigg),
	\end{align*}
where $n_{0v}$ denotes the minimum number of units in each terminal node. 	
\end{result}

\begin{proof}
\cite{dagdoug2020model} noted that the model-assisted estimator $\widehat{t}_{rf}$ given by (\ref{rf}) can also be viewed as a bagged estimator
\begin{equation*}
\widehat{t}_{rf}= \dfrac{1}{B} \sum_{b = 1}^B \widehat{t}_{tree}^{(b)},
\end{equation*}
where
$$\widehat{t}_{tree}^{(b)}=\displaystyle \sum_{k\in U}\widehat{f}^{(b)}_{tree}(\mathbf{x}_k, \theta_b)+\sum_{k\in S}\frac{y_k-\widehat{f}^{(b)}_{tree}(\mathbf{x}_k, \theta_b)}{\pi_k}$$
is a model-assisted estimator of $t_y$ based on the $b$-th  stochastic regression tree. Since
	\begin{align*}
\frac{1}{N_v}\mathbb{E}_p \bigg\rvert \widehat{t}_{rf} -t_y\bigg\rvert  \leqslant \dfrac{1}{B} \sum_{b = 1}^B  \mathbb{E}_p \bigg[   \bigg\rvert \dfrac{1}{
	N_v} \left(\widehat{t}_{tree}^{(b)} -t_y \right)\bigg\rvert  \bigg],
\end{align*}
the results follows from noting that the model-assisted estimator $\widehat{t}_{tree}^{(b)} $ is $L^1$ design-consistent for the total $t_y,$ $b=1, \cdots, B;$ see Result  \ref{res3}. 
\end{proof}

 \section{Simulation study} \label{sec5}

 In this section, we provide an empirical comparison of several model-assisted estimators. In addition to the estimators discussed in Sections 3 and 4, we have included  model-assisted estimators based on least angle regression \citep{efron2004least}, principal component regression, $k$-nearest neighbors, XGBoost \citep{chen2016} and Cubist \citep{quinlan1992learning}. For a description of these methods, see \cite{hastie_tibshirani_friedman_2011} and \cite{dagdoug2020imputation} and the references therein.

 We used data from the Irish Commission for Energy Regulation (CER) Smart Metering Project that was conducted in 2009-2010 (CER, 2011)\footnote{The data are available on request at: \texttt{https://www.ucd.ie/issda/data/commissionforenergyregulationcer/}. } \citep{cardot_goga_shehzad_2017}.
This  project  focused  on energy consumption and energy regulation. About 6000 smart meters were installed  to collect the electricity consumption of Irish residential and business customers every half an hour over a period of about two years.

 We considered a period of $14$ consecutive days and a population of $N=6,291$ smart meters (households and companies). Each day consisted of 48 measurements, leading to 672 measurements for each household. We denote by $X_j=X(t_j), j=1, \ldots, 672,$ the electricity consumption (in kW) at instant $t_j$ and by $x_{ij}$ the value of $X_j$ recorded by the $i$th smart meter  for $i=1, \ldots, 6,291.$  It should be noted that the matrix $N^{-1}\mathbf{X}^{\top}\mathbf{X}$  was ill-conditioned with a condition number equal to $254\ 753$. This suggests that some of the  $X$-variables were highly correlated with each other.

  We generated four survey variables based on these auxiliary variables according to the following models:
 \begin{align*}
 Y_1 &= 400+2X_1 +X_2 +2X_3 +\mathcal{N}(0, 1500); \\
 Y_2 &=500 + 2 X_4 + 400 \mathds{1} \left(X_5 > 156\right)- 400 \mathds{1} \left(X_5 \leqslant 156\right)+ 1000 \mathds{1} \left(X_2 > 190\right) \\&+300 \mathds{1} \left(X_5 > 200\right) + \mathcal{N}(0, 1500);\\
 Y_3 &= 1 + \cos(2X_1 + X_2 + 2X_3)^2 + \epsilon_1; \\
 Y_4 &= 4 +  3 \cdot \V \left(\left\{X_1 + X_{2}\right\}^2\right)^{-1/2} \times \left\{X_1 + X_{2}\right\}^2 + \mathcal{N}(0, 0.01),
 \end{align*}
where $\V(\cdot)$ denotes the empirical variance and the errors $\epsilon_1$ in the model for $Y_3$ were generated from an $\mathcal{E}xp(10)$  and these errors were centered so as to obtain a mean equal to zero.

Our goal was to estimate the population totals $t_{y_{j}}=\sum_{i \in U} y_{ji}, j=1, \cdots, 4.$   From the population, we selected  $R=2,500$ samples, of size $n = 600,$ which corresponds to a sampling fraction $n/N$ of about 10\%. We considered two sampling schemes: simple random sampling without replacement and stratified sampling with both optimal allocation and  proportional allocation.

 In each sample, we computed twelve model-assisted estimators of the form
 \begin{equation*}
 \widehat{t}_{ma}^{(j)} = \sum_{i \in U} \widehat{f}^{(j)}(\bx_i) + \sum_{i \in S} \dfrac{y_i - \widehat{f}^{(j)}(\bx_i)}{\pi_i}, \quad j =1, 2, \ldots, 12,
 \end{equation*}
 where the predictors $\widehat{f}^{(j)}(\bx_i)$, $j=1, 2, \ldots, 12,$ were obtained using the following procedures:
 \begin{enumerate}[Procedure 1:]
 	\item "LR" : Deterministic linear regression, leading to the GREG estimator.
 	\item "CART": Classification and regression tree algorithm \citep{breiman1984classification}, leading to an estimator closely related to that of \cite{mcconville2019automated} and implemented with the $R$-package \texttt{rpart}.
 	\item "RF": Random forests with the algorithm of \cite{breiman2001random} with $B=1000$ trees, a minimal number of elements in each terminal node $n_0 = 5$ and $p_0 = \lfloor\sqrt{p}\rfloor$, where $\lfloor \cdot\rfloor$ denotes the customary floor function. The algorithm leads to the estimator described in \cite{dagdoug2020model}. Simulations were implemented with the $R$-package \texttt{ranger}.
 	\item "Ridge": Ridge regression with a regularization parameter determined by cross-validation and  implemented with the $R$-package \texttt{glmnet}. The estimator was studied by \cite{goga2010overview}.
 	\item "Lasso": Lasso regression with a regularization parameter determined by cross-validation and  implemented with the $R$-package \texttt{glmnet} \citep{mcconville2017model}.
 	\item "EN": Elastic net regression with penalization coefficients determined by cross-validation with the $R$-package \texttt{glmnet}.
 	\item "XGB": XGBoost algorithm \citep{hastie_tibshirani_friedman_2011} with $50$ trees in the additive model, each tree being with a depth of at most $6$ and a learning rate $\lambda = 0.01$.  Simulations were implemented with the $R$-package \texttt{XGBoost}.
 	\item "5NN": $5$-nearest neighbors predictor with the euclidean distance  and implemented with the $R$-package \texttt{caret}.
 	\item "Cubist": A cubist algorithm \citep{kuhn_johnson} with $5$ models in each predictor, implemented with the $R$-package \texttt{cubist}; the algorithm and its adaptation for survey data are described in \cite{dagdoug2020imputation}.
 	\item "PCR1": Principal component regression based on the first $\lfloor p^{1/4} \rfloor$ components kept  and implemented with the $R$-package \texttt{pls} \citep{cardot_goga_shehzad_2017}.
 	\item "PCR2": Principal component regression based on the first $\lfloor p^{2/4} \rfloor$ components kept.
 	\item "PCR3": Principal component regression based on the first $\lfloor p^{3/4} \rfloor$ components kept.
 \end{enumerate}

As a measure of bias of the model-assisted estimators $\widehat{t}_{ma}^{(j)}$, $j=1, 2, ..., 12$, we computed the Monte Carlo percent relative bias defined as
 \begin{equation*}
 RB_{MC}\left(\widehat{t}_{ma}^{(j)}\right)=100 \times \dfrac{1}{R} \sum_{r=1}^R \dfrac{ (\widehat{t}_{ma}^{(j,r)} - t_y)}{t_y}, \quad j=1, 2, \ldots, 12,
 \end{equation*}
where $\widehat{t}_{ma}^{(j,r)}$ denotes the estimator  $\widehat{t}_{ma}^{(j)}$ at the $r$th iteration, $r=1, \ldots, R$. As a measure of efficiency, we computed the relative of efficiency, using the Horvitz-Thompson estimator $\widehat{t}_{\pi}$ given by (\ref{ht}), as the reference. That is,
 \begin{equation*}
 RE_{MC}\left(\widehat{t}_{ma}^{(j)} \right)=100 \times \dfrac{ MSE_{MC}(\widehat{t}_{ma}^{(j)}) }{ MSE_{MC}(\widehat{t}_{\pi})},  \quad j=1, 2, ..., 12,
 \end{equation*}
 where $MSE_{MC}( \widehat{t}_{ma}^{(j)} ) = R^{-1} \sum_{r=1}^R ( \widehat{t}_{ma}^{(j, r)} - t_y )^2$ and $MSE_{MC}(\widehat{t}_{\pi})$ is defined similarly.

We were also interested in investigating to which extent the model-assisted estimators $\widehat{t}_{ma}^{(j)}, j=1, \ldots, 12$ were affected by the inclusion of a large number of predictors in the working models.  To that end, in addition to the variables $X_1$-$X_3,$ we included $d_{noise}$ predictors in the working models. These predictors were available in the Irish data set. We used the following values for $d_{noise}$:  5, 10, 20, 50, 100, 200, 300 and 400.



\subsection{Simple random sampling without replacement}

In this section, we present the results obtained under simple random sampling without replacement (SRSWOR) of size $n=600$. All the point estimators $\widehat{t}_{ma}^{(j)}, j=1, \ldots, 12,$ exhibited a negligible or small percent RB with a maximum value of about   $3.1 \%$ (obtained in the case of the GREG estimator). For this reason, results pertaining to relative bias are not reported here. 

\begin{figure}[h!]
	\centering
	\includegraphics[width=1\linewidth]{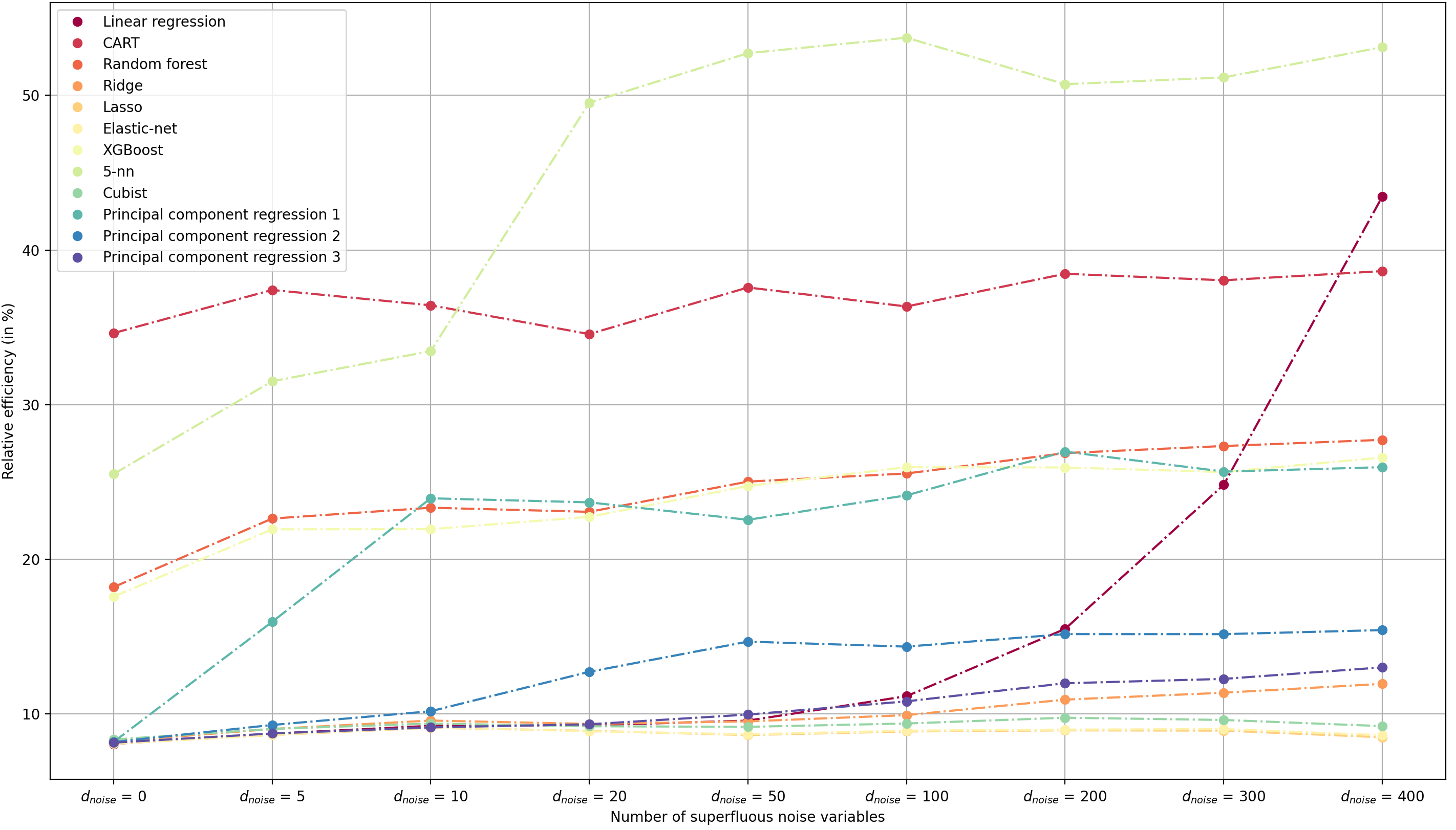}
	\caption{Relative efficiency of model-assisted estimators $\widehat{t}_{ma}^{(j)}, j=1, \ldots, 12$ for the estimation of the total of $Y_1$ with SRSWOR ($n=600$) and increasing number of auxiliary variables}
	\label{fig1}
\end{figure}

\begin{figure}[h!]
	\centering
	\includegraphics[width=1\linewidth]{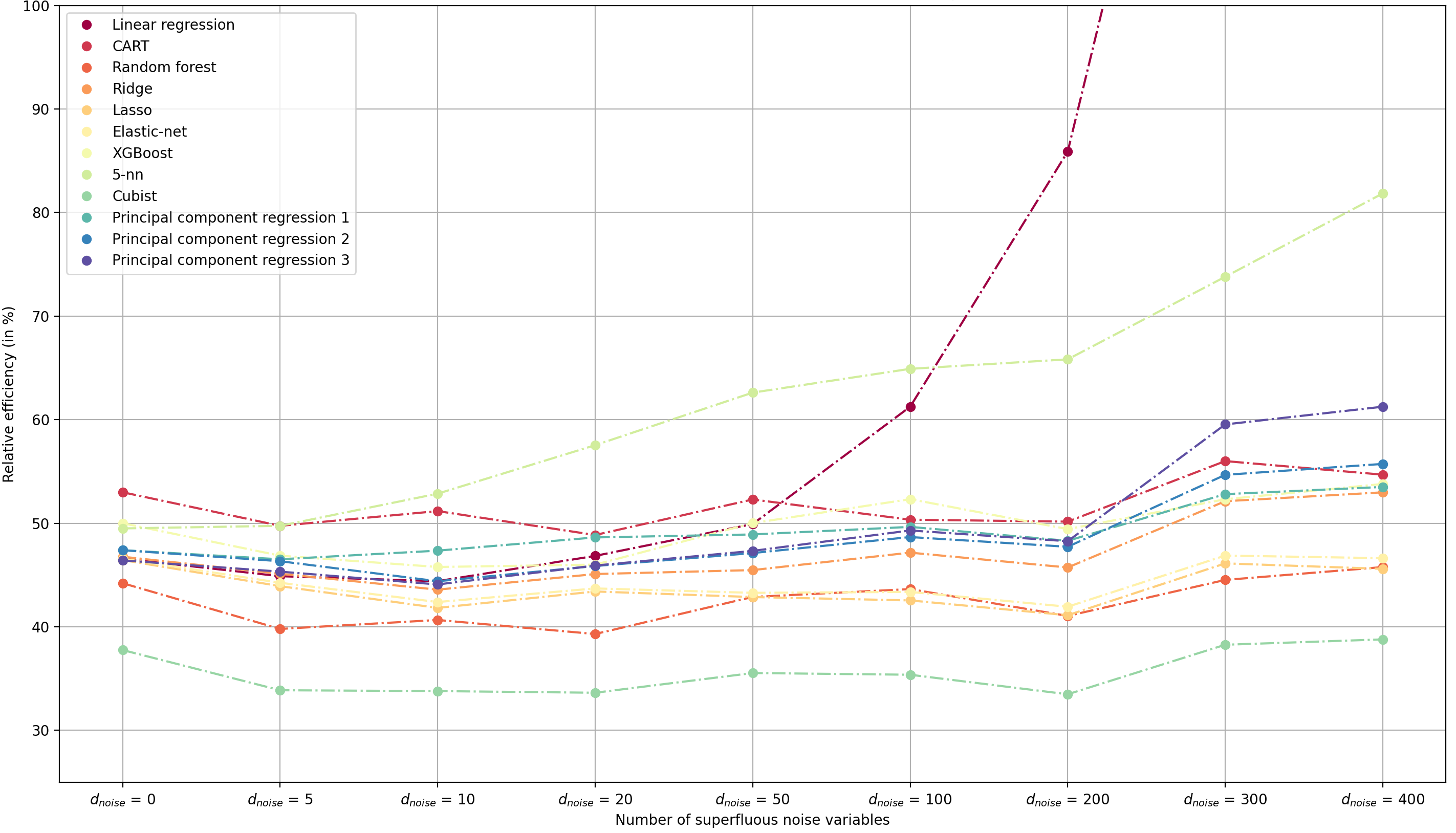}
	\caption{Relative efficiency  of model-assisted estimators $\widehat{t}_{ma}^{(j)}, j=1, \ldots, 12$  for the estimation of the total of $Y_2$ with SRSWOR, $n=600$ and increasing number of auxiliary variables}
	\label{fig2}
\end{figure}

\begin{figure}[h!]
	\centering
	\includegraphics[width=1\linewidth]{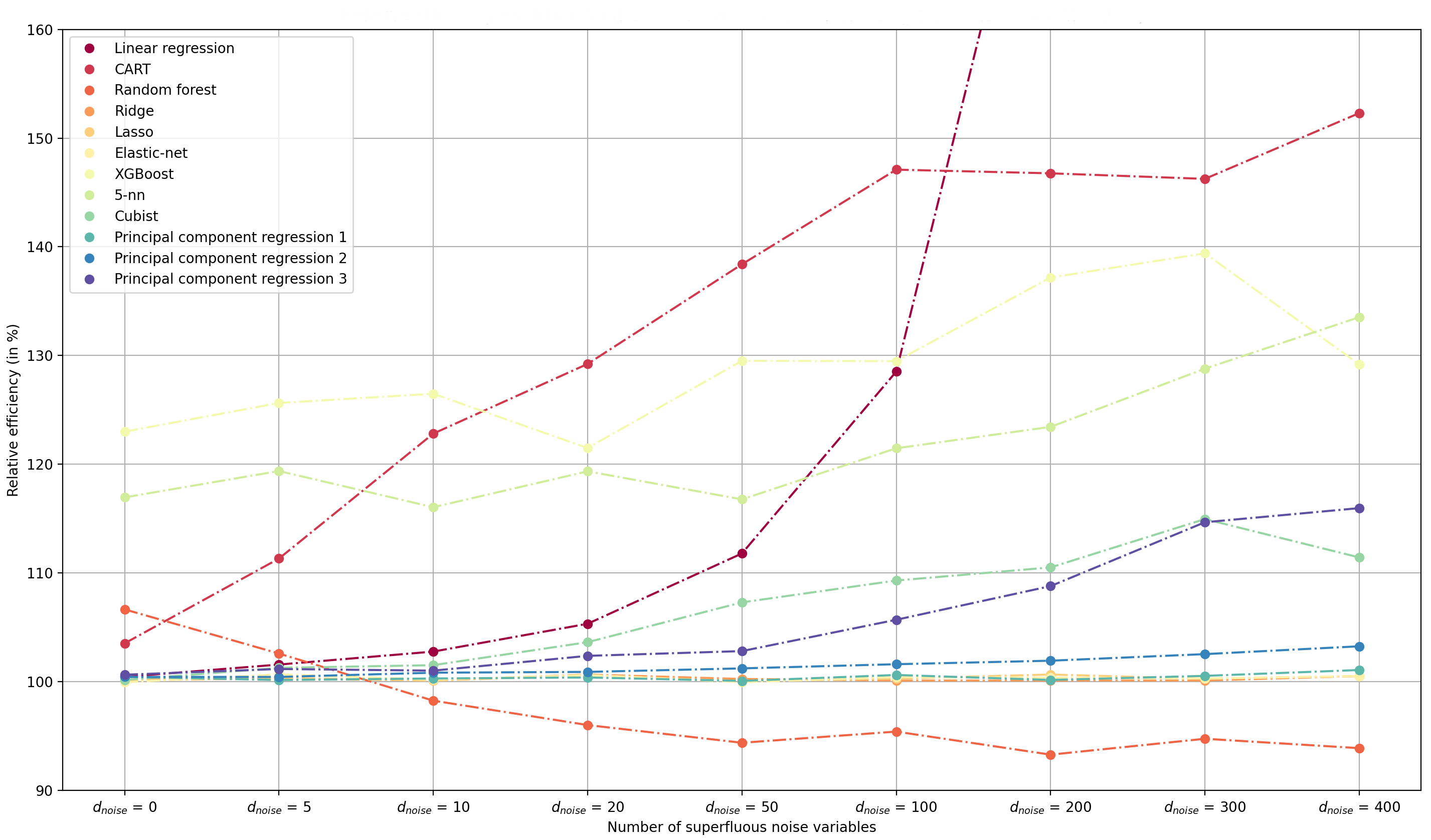}
	\caption{Relative efficiency  of model-assisted estimators $\widehat{t}_{ma}^{(j)}, j=1, \ldots, 12$ for the estimation of the total of $Y_3$ with SRSWOR, $n=600$ and increasing number of auxiliary variables}
	\label{fig3}
\end{figure}

\begin{figure}[h!]
	\centering
	\includegraphics[width=1\linewidth]{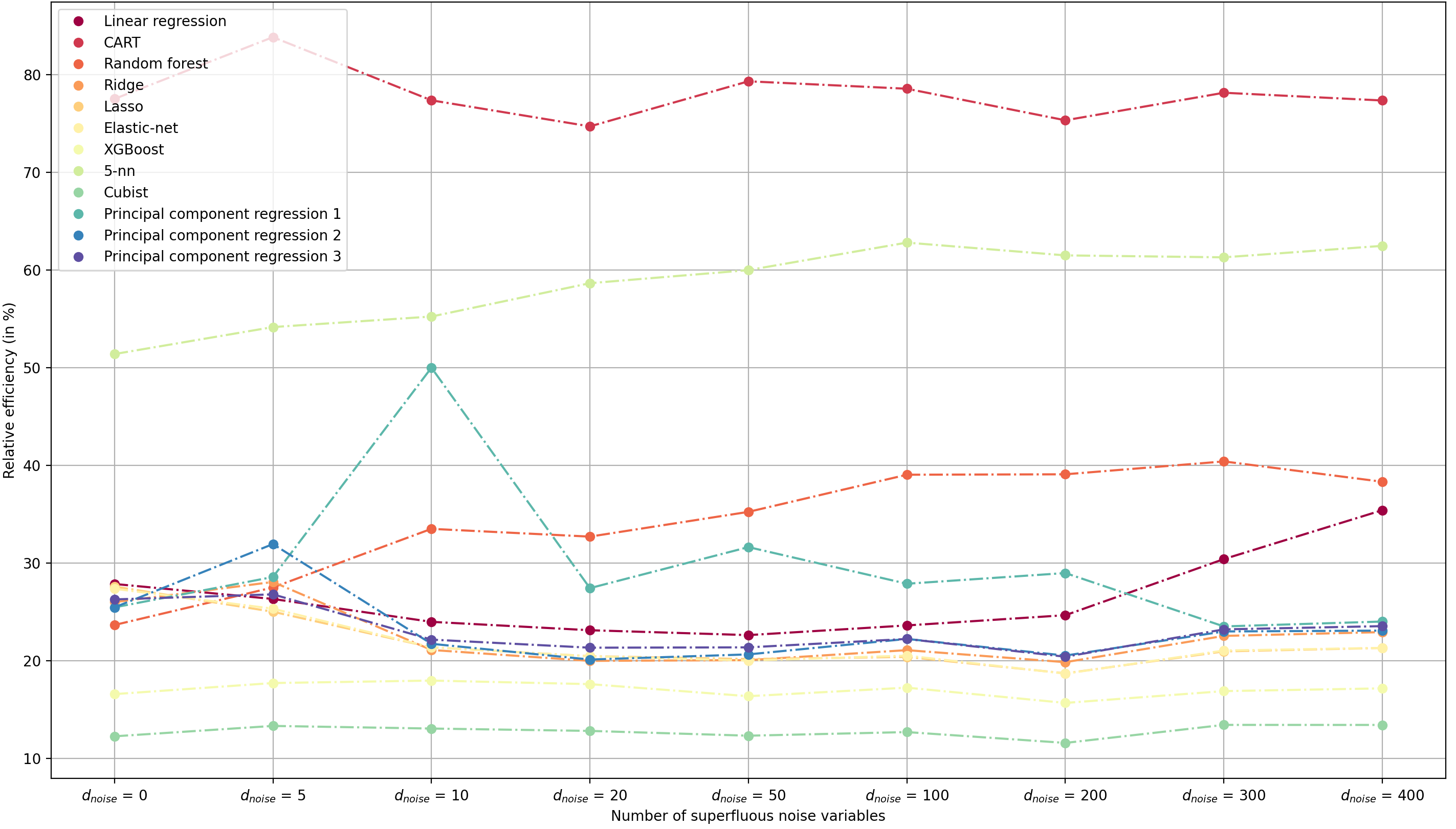}
	\caption{Relative efficiency  of model-assisted estimators $\widehat{t}_{ma}^{(j)}, j=1, \ldots, 12$ for the estimation of the total of $Y_4$ with SRSWOR, $n=600$ and increasing number of auxiliary variables}
	\label{fig4}
\end{figure}

Figures \ref{fig1}-\ref{fig4} display the relative efficiency of the model-assisted estimators $\widehat{t}_{ma}^{(j)}, j=1, \ldots, 12$ as a function of the number of auxiliary variables incorporated in the working models. To improve readability, we have truncated some large values of RE, when applicable.


We begin by discussing the results on relative efficiency pertaining to the estimation of the total of the survey variable $Y_1$. For low-dimensional settings, the GREG estimator was very efficient with values of RE below $10\%$.  These results can be explained by the fact that $Y_1$ was linearly related to the $\mathbf{x}$-variables. However, as the number of variables $d_{noise}$ increased, the efficiency of the GREG estimator rapidly deteriorated, suggesting that the performance of the GREG estimator is sensitive to the dimension of the $\mathbf{x}$-vector.
As expected, model-assisted estimators based on regularization methods such as ridge, lasso, elastic-net or dimension reduction methods such as principal components regression, performed generally very well. Unlike the GREG, these estimators were not much affected by the number of auxiliary variables incorporated in the model.   Turning to the model-assisted estimator based on a $5$-nn, we note that it was less efficient than most competitors and that its efficiency got worse as $d_{noise}$ increased, a phenomenon referred to as the curse of dimensionality. The model-estimators based on  XGBoost, Cubist and random forests performed quite well and did not seem to be affected by the number of auxiliary variables incorporated in the model. Finally, the estimators based on CART were less efficient than those obtained through the other machine learning methods.

The results pertaining to the survey variable $Y_2$ and displayed in Figure \ref{fig2} were fairly consistent with those obtained for the survey variable $Y_1$ with one exception: the Cubist algorithm was significantly more efficient than the other procedures in  all the scenarios.

Turning to the survey variable $Y_3$ (see Figure \ref{fig3}), the model-assisted estimator based on random forests was significantly more efficient than the Horvitz--Thompson estimator, especially for large values of  $d_{noise}$. The other procedures led to estimators less efficient than the Horvitz-Thompson estimator with values of RE above $100$.  In particular, the GREG estimator broke down as the number of auxiliary variable increased. The performance of model-assisted estimators based on CART and XGBoost algorithms deteriorated as the dimension  increased. In a high-dimension setting with highly correlated predictors, random forests improved over CART due  to the random subsampling of $p_0$ variables among the $p$ variables, generating then decorrelated trees \citep{hastie_tibshirani_friedman_2011}.

The results in Figure \ref{fig4} about the survey variable $Y_4$ were similar to the ones in previous figures. Most estimators remained mostly unaffected by the number of auxiliary variables  $d_{noise}$. Again, the model-assisted estimator based on the Cubist algorithm was the best in all the scenarios.


\subsection{Stratified simple random sampling with optimal allocation}

In the second simulation study, we partitioned  the Irish residential and business customer population into four strata $U_1, \ldots, U_4,$  using an equal quantile method with respect to the variable, $X_1,$ the electricity consumption at instant $t_1.$  From the population, we selected $R=2,500$ stratified simple random samples, of size $n=600.$ The stratum sample sizes $n_h$ were determined using an $X_2$-optimal allocation, where $X_2$ denotes the electricity consumption recorded at instant $t_2.$ This led to $n_1=20, n_2=36, n_3=45$ and $n_4= 499.$  The first-order inclusion probabilities, $\pi_i=n_h/N_h, i \in U_h$ and the sampling weights $w_i=\pi_i^{-1}$ are shown in Table \ref{tab:STrat}.

\begin{table}[]
	\begin{center}
	\begin{tabular}{ccccc}
		Stratum        & 1     & 2     & 3     & 4     \\ \hline
		$\pi_i$        & 0.012 & 0.022 & 0.028 & 0.316 \\ \hline
		$w_i=\pi_i^{-1}$ &   77.85    &    43.83   &   35.11    &  3.16
	\end{tabular}
	\caption{First-order inclusion probabilities and sampling weights within strata.}
\label{tab:STrat}
\end{center}
\end{table}
We confined to the survey variables $Y_1$ and $Y_3$ only and we aimed at estimating $t_{y_1}$ and $t_{y_3}$. It is worth pointing out that the resulting sampling design was informative as the variables used at the design stage ($X_1$ and $X_2$) were also related to the survey variables $Y_1$ and $Y_3$.  In fact, the Monte Carlo coefficient of correlation between the sampling weights and $Y_1$ was approximately equal to $0.402.$ We do not report the coefficient of correlation between between the sampling weights and $Y_3$ as the relationship between $Y_3$ and the set of predictors $X_1, X_3$ is not linear.

Again, in each sample we computed twelve model-assisted estimators  $\widehat{t}_{ma}^{(j)}, j=1, \ldots, 12$ for each of $t_{y_1}$ and $t_{y_3}.$ Since most machine learning software packages do not take the sampling weights into account, we have included the design variables $X_1$ and $X_2$ in the set of predictors.

\begin{figure}[h!]
	\centering
	\includegraphics[width=1\linewidth]{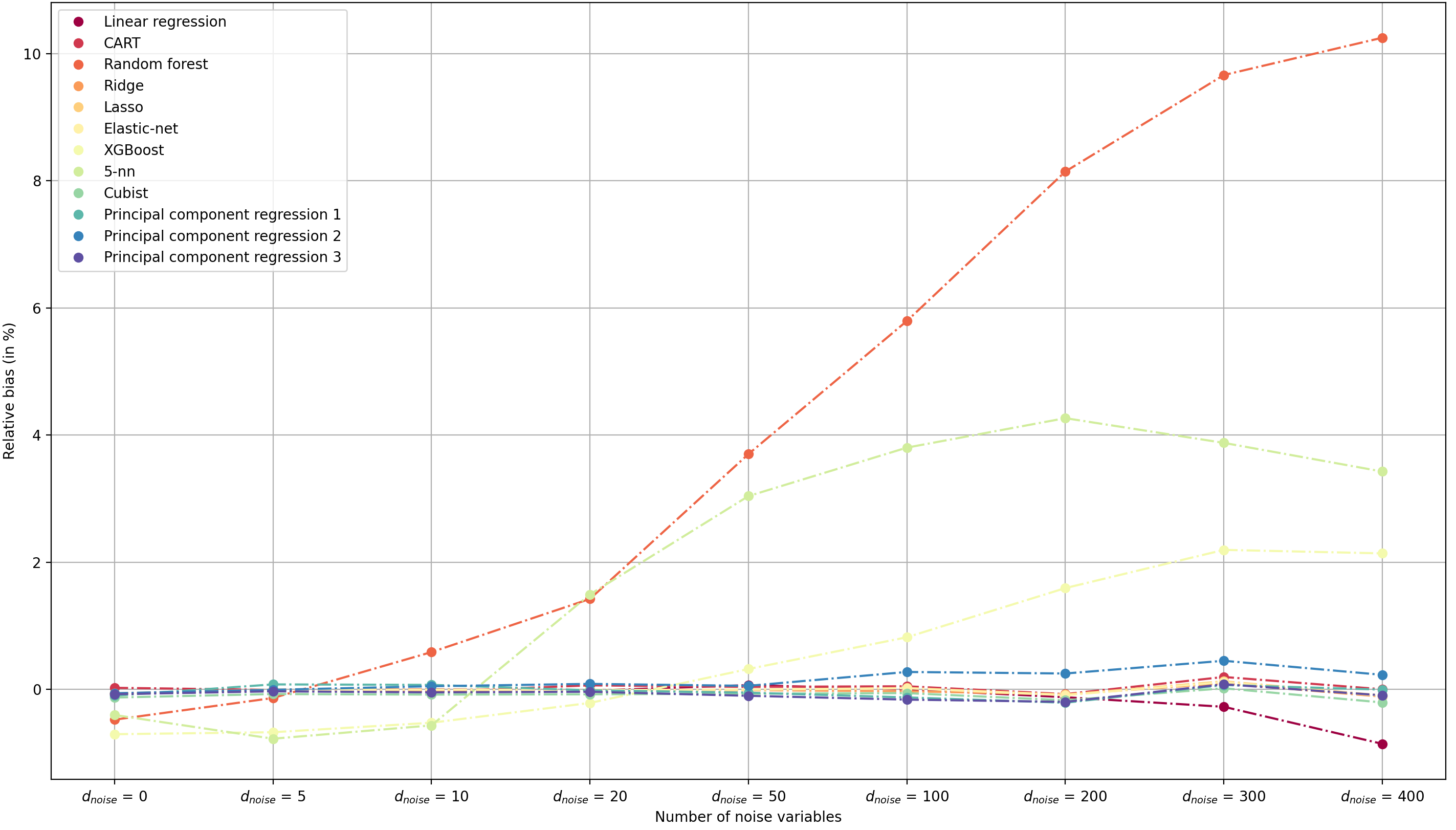}
	\caption{Relative bias of model-assisted estimators $\widehat{t}_{ma}^{(j)}, j=1, \ldots, 12$ for the estimation of $Y_1$ with stratified sampling with $X_2$-optimal allocation, $n=600$ with increasing number of auxiliary variables}
	\label{fig6}
\end{figure}

\begin{figure}[h!]
	\centering
	\includegraphics[width=1\linewidth]{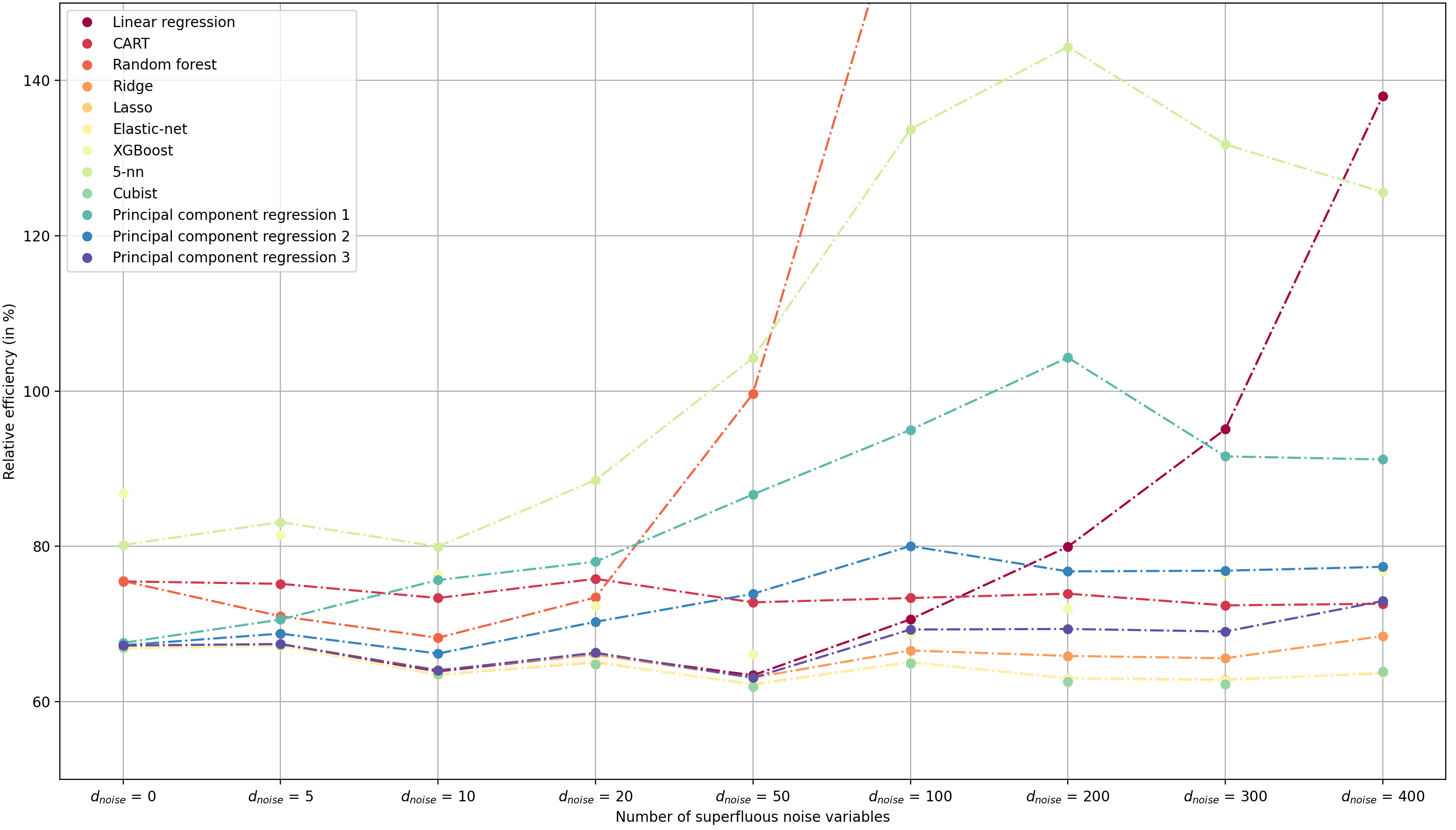}
	\caption{Relative efficiency  of model-assisted estimators $\widehat{t}_{ma}^{(j)}, j=1, \ldots, 12$ for the estimation of the total of $Y_1$ with stratified sampling with $X_2$-optimal allocation, $n=600$ and increasing number of auxiliary variables}
	\label{fig7}
\end{figure}

We begin by discussing the results pertaining to the estimation of the total of the survey variable $Y_1.$ Figure \ref{fig6} and Figure \ref{fig7} display the Monte Carlo percent relative bias and the Monte Carlo relative efficiency as a function of the number of variables $d_{noise}$.  Except for the model-assisted estimators based on $5$-nn and random forest, the other estimators exhibit a small value of RB for all values of $d_{noise}$. Again, the $5$-nn model-assisted estimator suffered from the curse of dimensionality. Turning to the estimator based on random forests, we note from Figure  \ref{fig6} that the bias increased as the number of predictors  $d_{noise}$ increased. For instance, for $d_{noise}=400,$ the value of RB was just above 10\%. This significant bias may be explained by the fact that random forests is the only procedure among the ones considered in our simulation that randomly selects $p_0=\sqrt{p}$ variables among the initial $p$ predictors at each split. For instance, for $d_{noise}=400$, only 20 variables are randomly selected at each split.  As a result, most predictions obtained trough a random forests algorithm were based on misspecified working models, leading to potentially bad fits and large residuals. Also, each prediction corresponds to a weighted mean computed within each node with $n_0=5$ observations only. Therefore, each predictions corresponds to a ratio-type estimate based on 5 observations only. This, together with the fact that the sampling weights are highly variable, constitutes a conducive ground for the occurrence of small sample bias. In terms of efficiency, except for the GREG, the $5$-nn and the random forest estimators, the other procedures performed well with values of RE ranging from 60\% to 80\%. The best procedures were Cubist and Lasso.

\begin{figure}[h!]
	\centering
	\includegraphics[width=1\linewidth]{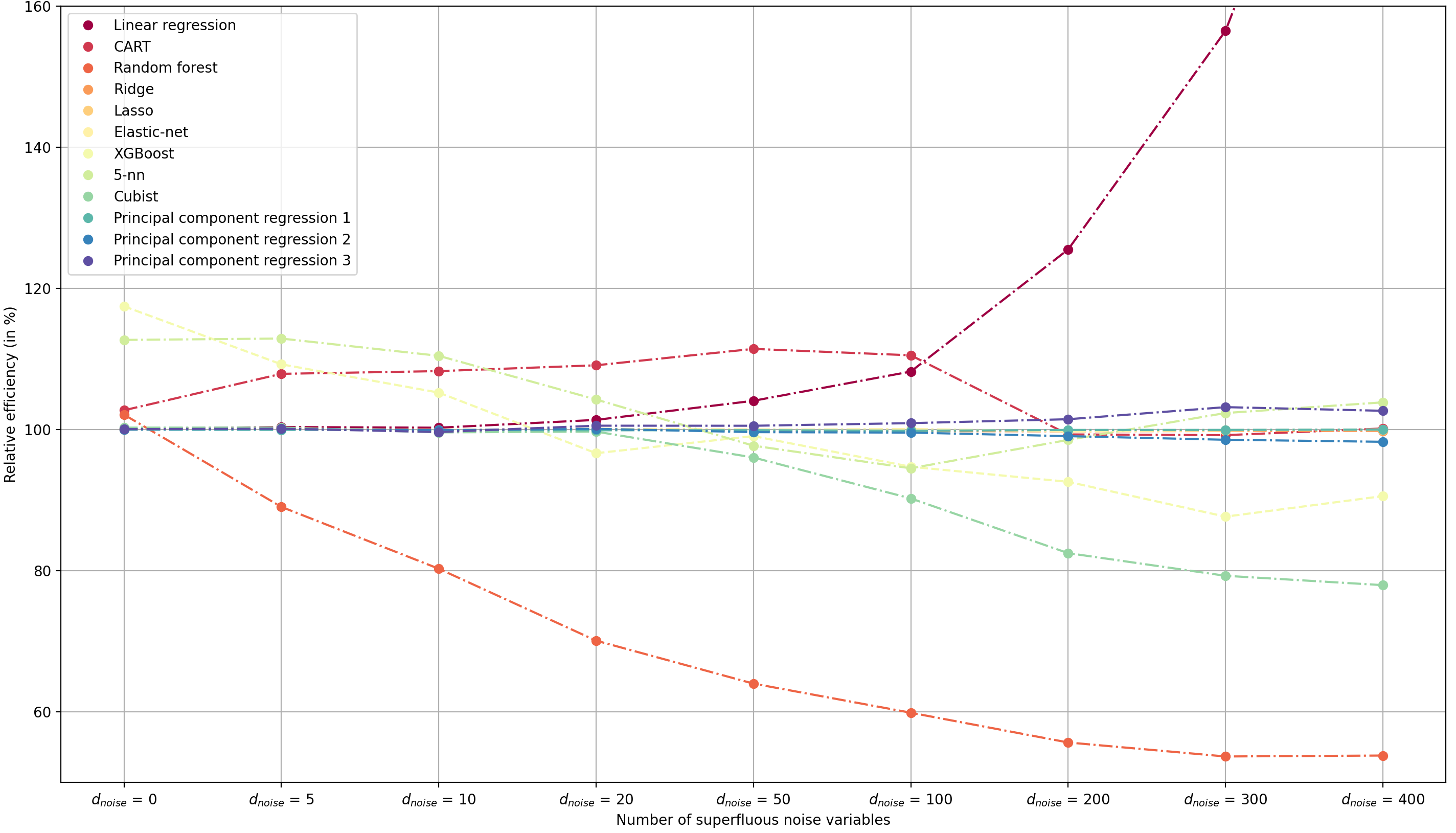}
	\caption{Relative efficiency  of model-assisted estimators $\widehat{t}_{ma}^{(j)}, j=1, \ldots, 12$ for the estimation of the total of $Y_3$ with stratified sampling with $X_2$-optimal allocation, n=600 and increasing number of auxiliary variables}
	\label{fig5}
\end{figure}

We now turn to the survey variable $Y_3.$ First, the Monte Carlo relative bias was negligible for all the estimation procedures and are not reported here. Results about relative efficiency are  plotted in Figure \ref{fig5}. Random forests performed extremely well and their performance improved as $d_{noise}$ increased. This suggests that the method was able to extract the information contained in the predictors. This was also true for Cubist and XGBoost, although to a lesser extent.

\begin{figure}[h!]
	\centering
	\includegraphics[width=1\linewidth]{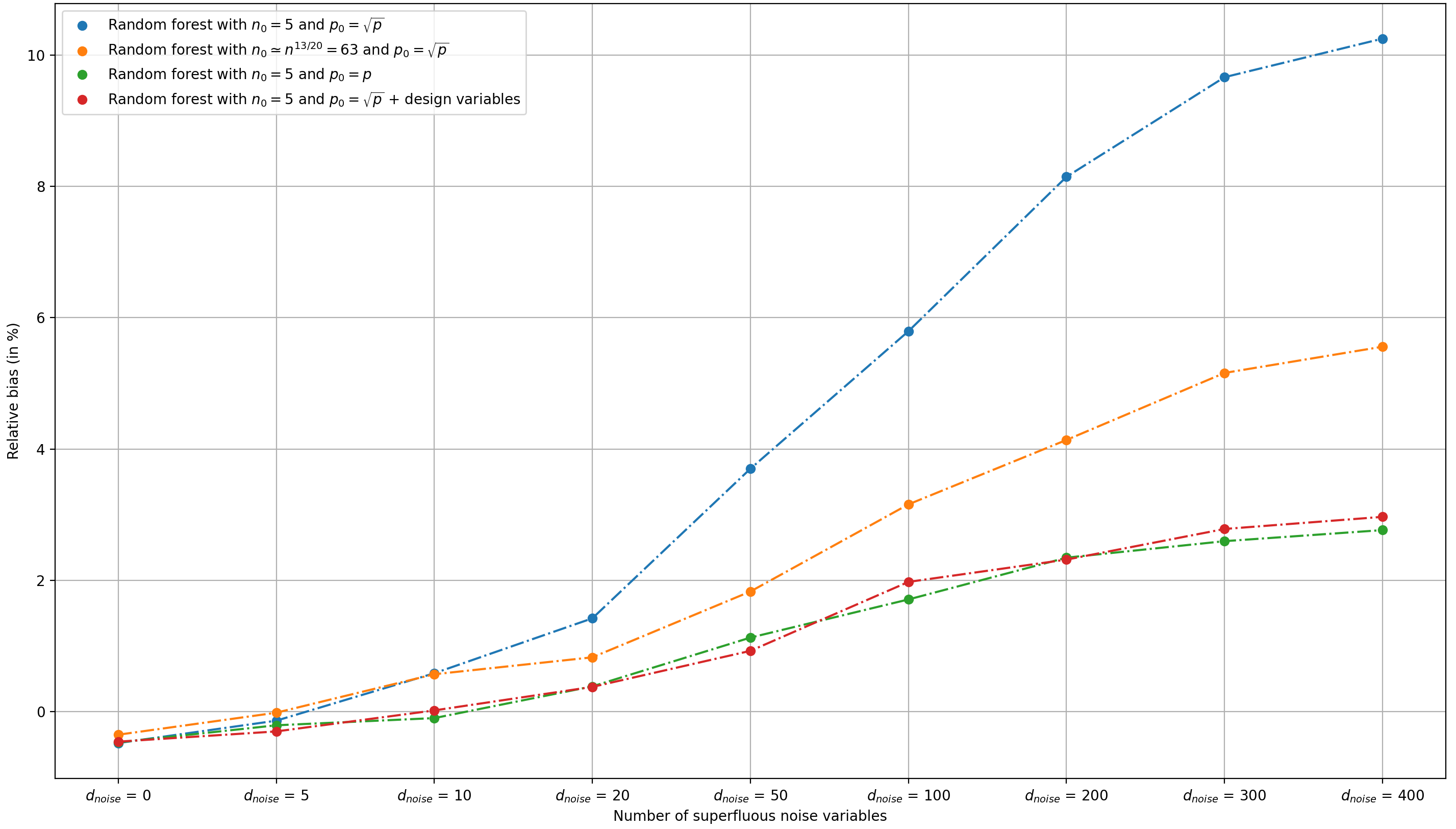}
	\caption{Comparison of different configurations of hyper-parameters for $\widehat{t}_{rf}$ for the estimation of the total of $Y_1$ with stratified sampling and $X_2$-optimal allocation, $n=600$.}
	\label{fig8}
\end{figure}

To get a better understanding of the performance of random forests for the survey variable $Y_1,$ we conducted additional scenarios based on different values of the hyper parameters $n_0,$ the number of observations within each terminal nodes, and  $p_0,$ the number of variables randomly selected at each split among the initial $p$ model variables. We used the following values for $n_0$ and $p_0$:
\begin{itemize}
\item $n_0=5$ observations and $p_0=\sqrt{p}$ variables which are the default choices in the $R$-package \texttt{rangers};
\item $n_0=5$ observations and $p_0=p$ variables;
\item $n_0=5$ observations and $p_0=\sqrt{p}$ variables and the stratification variable $X_1$ was selected with probability 1, at each split, besides the $p_0$ variables;
\item $n_0=n^{13/20}$ observations and $p_0=\sqrt{p}$ variables.
\end{itemize}
The Monte Carlo percent relative bias is displayed in  Figure \ref{fig8}. We note that  relative bias was much smaller when all the predictors were considered at each split or if the stratification variables were considered besides  $p_0$ variables at each split. Moreover, the bias decreased when more observations were allowed  in each terminal node. These results suggest, that, in order to avoid significant small sample bias,  we recommend to use more observations in each terminal nodes and/or to avoid the randomization in the process of growing trees and/or to force the design variables to be selected at each split. The latter option is available in the $R$ package \texttt{ranger}. 

\subsection{Stratified sampling with proportional allocation}

\begin{figure}[h!]
	\centering
	\includegraphics[width=0.8\linewidth]{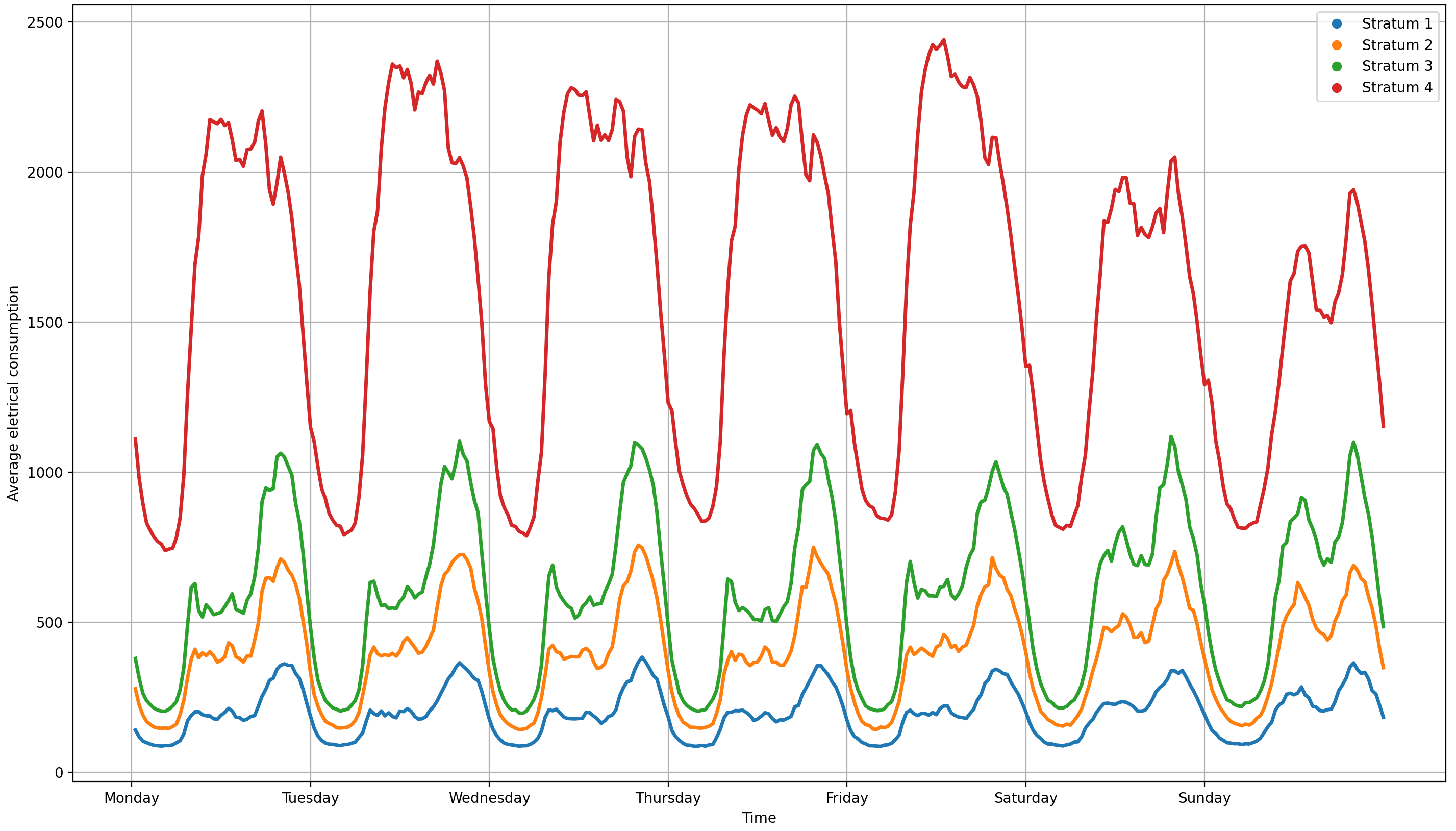}
	\caption{Average electricity consumption on each stratum during first week}
	\label{fig9}
\end{figure}

In this section, we consider a more realistic scenario based again on the Irish residential and business customer data. As a stratification variable, we used the mean electricity consumption recorded during the first week. Again, we constructed four strata using an equal-quantile method based, this time, on the mean electricity consumption; see also \cite{CDGJL2012} who used a similar design.
The mean trajectories during the first week within each stratum are  plotted in Figure \ref{fig9}. From Figure \ref{fig9}, we note that Stratum 1 corresponds to consumers with low global levels of electricity consumption, whereas Stratum 4 consists of consumers who have high levels of electricity consumption.

Our aim was to estimate the total electricity consumption recorded on the Monday of the second week and given by $t_y=\sum_{i=1}^{6291}\sum_{j=336}^{384}y_{ij},$ where $y_{ij}$ is the electricity consumption recorded for the $i$-th unit at the $j$-th instant. Within each stratum, we selected a sample, of size $n_h,$ according to simple random sampling without replacement. The $n_h$'s were determined according to proportional allocation; i.e, $n_h=n \times (N_h/N)$ with $n=600$. In each of the 2,500 samples, we computed the same 12 model-assisted estimators as in the previous sections. Again,  we computed the Monte Carlo percent relative bias and the relative efficiency for each the 12 estimators. The results are presented in Table \ref{tab:1}.


\begin{table}[h!]
	\centering
	\begin{tabular}{ccccc}
		\hline
		Estimator            &                      & Relative bias        &                      & Relative efficiency  \\ \hline
		&                      &                      &                      &                      \\
		LR                   &                      & 0.2                 &                      & 9.3                 \\
		CART                 &                      & -0.1                &                      & 41.0                \\
		RF                   &                      & -1.1                &                      & 17.0                \\
		Ridge                &                      & 0.1                 &                      & 4.0                 \\
		Lasso                &                      & 0.2                 &                      & 4.1                 \\
		EN                   &                      & 0.2                 &                      & 4.1                 \\
		XGB                  &                      & -1.7                &                      & 24.9                \\
		NN5                  &                      & -4.0                &                      & 65.6                \\
		Cubist               &                      & -0.0                &                      & 4.3                 \\
		PCR1                 &                      & 0.1                 &                      & 4.9                 \\
		PCR2                 &                      & 0.1                 &                      & 4.2                \\
		PCR3                 &                      & 0.1                 &                      & 4.2                 \\
		\multicolumn{1}{l}{} & \multicolumn{1}{l}{} & \multicolumn{1}{l}{} & \multicolumn{1}{l}{} & \multicolumn{1}{l}{} \\ \hline
	\end{tabular}
	\caption{Monte Carlo percent relative bias and relative efficiency of several model-assisted estimators under stratified simple random sampling with proportional allocation.}
	\label{tab:1}
\end{table}
From Table \ref{tab:1}, we note that the $5$-nn model-assisted estimator was the only estimator to exhibit a non-negligible bias. Although it was less efficient than its competitors, it was more efficient than the Horvitz-Thompson estimator with a value of RE of about $65.6\%$. The ridge estimator was the most efficient with a value of RE equal to $4\%$ and was closely followed by lasso, elastic-net, Cubist and principal components model-assisted estimators. The GREG estimator performed very well with a value of RE of about 9.3\%.  Random forests led to considerable improvement over the CART model-assisted estimator with values of RE of 17\% and 41\%, respectively. Still,  random forests were less efficient than the GREG estimator, which is not surprising as the relationship between the survey variable and the auxiliary variables was linear.

\section{Final remarks} \label{sec6}
In this paper, we have examined a number of model-assisted estimation procedures in a high-dimensional setting both theoretically and empirically.  If the relationship between the survey variable and the auxiliary information can be well described by a linear model, our results suggest that penalized estimators such as ridge, lasso and elastic net perform very well in terms of bias and efficiency, even in the case $p=n$. Model-assisted estimators based on random forests, Cubist and XGBoost methods were mostly unaffected by the number of predictors incorporated in the working model, even in the case of complex relationships between the study and the auxiliary variables. As expected, the GREG estimator suffered from poor performances in the case of a large number of auxiliary variables.

The procedure Cubist stood out from the other machine learning procedure with very good performances in virtually all the scenarios. Further work is needed to establish the theoretical properties of model-assisted estimators based on Cubist in both a low-dimensional and high-dimensional settings.

Variance estimation is an important stage of the estimation process. Further research includes  identifying the regularity conditions under which the variance estimators are design-consistent in a high-dimensional setting.

We end this article by mentioning that virtually all the machine learning software packages cannot handle design features such as unequal weights and stratification. For instance, some random forests algorithms may involve a bootstrapping procedure and/or a cross-validation procedure. To fully account for the sampling design, both procedures must be modified so as to account for the design features. Not fully accounting for the sampling design may be viewed as a form of model misspecification. However, model-assisted estimation procedures remain design-consistent even if the model is misspecified.  In our experiments, several machine learning procedures (e.g., random forests, Cubist, XGboost) performed very well in most scenarios even though we did not modify the bootstrapping and cross-validation procedures to account for design features.  In other words, it seems that, accounting for predictors that are highly predictive of the $Y$-variable, seems to be the preponderant factor with respect the to efficiency aspect of model-assisted estimators. We conjecture that, fully accounting for the sampling design, will likely lead to additional efficiency gains but that the predictive power of the model likely constitutes the "determining factor". Developing machine learning procedures that fully account for the sampling design is currently under investigation.

\section*{Acknowledgment}
We thank the Editor, an Associate Editor, and two referees for their comments and suggestions, which helped improving the paper substantially.  We are very grateful to Professor Patrick Tardivel from the Université de Bourgogne for enlightening discussions about the lasso method.  The work of Mehdi Dagdoug was supported by grants of the Franche-Comté region and Médiamétrie. The work of David Haziza was supported by a grant of the Natural Sciences and Engineering Research Council of Canada.

 \bibliographystyle{apalike}
 \bibliography{biblio}

 \addcontentsline{toc}{section}{References}


\newpage

 \section*{Appendix}
 \subsection*{Proof of Result \ref{res1} }
             	We adapt the  proof of \cite{robinson_sarndal_1983} to a high-dimensional setting. Let $I_i$ be the sample membership indicator for unit $i$ such that $I_i=1$ if $i \in S$ and $I_i=0,$  otherwise. Let $\alpha_i := I_i / \pi_i -1$ for all $i\in U_v$. We consider the following decomposition:
	\begin{align} \label{decompo}
	\dfrac{1}{N_v}	\left( \widehat{t}_{\rm{greg}} -t_y \right)= \dfrac{1}{N_v} \sum_{i \in U_v}  \alpha_iy_i - \sum_{j=1}^{p_v}  b_j\widehat{\beta}_j,
	\end{align}
	where  $b_j= \dfrac{1}{N_v} \sum_{i \in U_v} \alpha_ix_{ij}$ for $j=1, 2, ..., p_v.$
Now, the first term does not depend on the auxiliary information and we have \citep{robinson_sarndal_1983, breidt_opsomer_2000}:
\begin{align} \label{a_v}
	 \mathbb{E}_{\rm{p}}\left(\dfrac{1}{N_v} \sum_{i \in U_v}  \alpha_iy_i\right)^2 &= \dfrac{1}{N_v^2} \sum_{i \in U} y_i^2 \cdot \mathbb{E}_{\rm{p}} (\alpha_i^2) + \dfrac{1}{N_v^2}\sum_{i\in U_v} \sum_{\ell\in U_v,\ell\neq i} y_iy_{\ell} \cdot \mathbb{E}_{\rm{p}} (\alpha_i \alpha_{\ell}).
	\end{align}
We have $\mathbb{E}_{\rm{p}} (\alpha_i^2)=(1-\pi_i)/\pi_i\leqslant 1/c$ and for $i\neq \ell,$ $\mathbb{E}_{\rm{p}} (\alpha_i \alpha_{\ell})=(\pi_{i\ell}-\pi_i\pi_{\ell})/\pi_i\pi_{\ell}\leqslant \max_{i,\ell\in U_v,i\neq \ell}|\pi_{i\ell}-\pi_i\pi_{\ell}|/c^2$ by Assumption (H\ref{H3}). It follows from (H\ref{H1}), (H\ref{H2}) and (H\ref{H3}) that
\begin{align}
\mathbb{E}_{\rm{p}}\left(\dfrac{1}{N_v} \sum_{i \in U_v}  \alpha_iy_i\right)^2&\leqslant \frac{1}{c N_v^2}\sum_{i\in U_v}y^2_i+\frac{n_v\max_{i,\ell\in U_v,i\neq \ell}|\pi_{i\ell}-\pi_i\pi_{\ell}|}{c^2 n_v N_v^2}\sum_{i\in U_v} \sum_{\ell\in U_v,\ell\neq i} |y_iy_{\ell}|\nonumber\\
&\leqslant \left(\frac{1}{cN_v}+\frac{n_v\max_{i,\ell\in U_v,i\neq \ell}|\pi_{i\ell}-\pi_i\pi_{\ell}|}{c^2 n_v}\right)\frac{1}{N_v}\sum_{i\in U_v}y^2_i=\mathcal{O}\left(\frac{1}{n_v}\right)\label{borne_a_v2}
\end{align}
and so,
\begin{eqnarray}
\bigg\rvert \dfrac{1}{N_v} \sum_{i \in U_v}  \alpha_iy_i\bigg\rvert =\mathcal{O}_{\rm p}\left(\frac{1}{\sqrt{n}_v}\right).\label{terme_A}
\end{eqnarray}
Now, consider the second term from the right-side of (\ref{decompo}):
\begin{eqnarray}
\bigg\rvert \sum_{j=1}^{p_v} \widehat{\beta}_j b_j\bigg\rvert \leqslant\sqrt{\left(\sum_{j=1}^{p_v} \widehat{\beta}_j^2\right)\left(\sum_{j=1}^{p_v}  b_j^2\right)}=||\widehat{\boldsymbol{\beta}}||_2\sqrt{\sum_{j=1}^{p_v}  b_j^2}\leqslant
||\widehat{\boldsymbol{\beta}}||_1\sqrt{\sum_{j=1}^{p_v}  b_j^2}.\label{terme_B}
\end{eqnarray}
By Assumption (H\ref{H4}), we have that $||\widehat{\boldsymbol{\beta}}||_1=\mathcal{O}_{\rm p}(p_v).$ Furthermore,
$$
\sqrt{\sum_{j=1}^{p_v}  b_j^2}=\frac{1}{N_v}\left|\left|\sum_{i\in U_v}\alpha_i\bx_i\right|\right|_2
$$
and
	\begin{align}
	 \frac{1}{N_v^2}\mathbb{E}_{\rm{p}}\left|\left|\sum_{i\in U_v}\alpha_i\bx_i\right|\right|_2^2&=  \dfrac{1}{N_v^2} \sum_{i  \in U_v} ||\bx_{i}||_2^2\mathbb{E}_{\rm{p}} (\alpha_i^2) +\dfrac{1}{N_v^2} \sum_{i\in U_v}\sum_{\ell\neq i  \in U_v} \bx^{\top}_{i} \bx_{\ell }\mathbb{E}_{\rm{p}} (\alpha_i \alpha_{\ell}) \nonumber\\
	 &\leqslant\dfrac{1}{c N_v^2}\sum_{i  \in U_v} ||\bx_{i}||_2^2+\dfrac{n_v\max_{i,\ell\in U_v,i\neq \ell}|\pi_{i\ell}-\pi_i\pi_{\ell}|}{c^2 n_vN_v^2}\sum_{i\in U_v}\sum_{\ell\neq i  \in U_v} | \bx^{\top}_{i} \bx_{\ell }|
	 \nonumber\\
	&\leqslant\left(\frac{1}{cN_v}+\frac{n_v\max_{i,\ell\in U_v,i\neq \ell}|\pi_{i\ell}-\pi_i\pi_{\ell}|}{c^2 n_v}\right)\frac{1}{N_v}\sum_{i\in U_v}||\bx_i||_2^2\nonumber\\
	&=  \mathcal{O}\bigg(\dfrac{p_v}{n_v}\bigg),\label{borne_x}
	\end{align}
by Assumptions (H\ref{H2})-(H\ref{Hnew}).  It follows that
\begin{eqnarray}
\sqrt{\sum_{j=1}^{p_v}  b_j^2}=\mathcal O_{\rm{p}}\bigg(\sqrt{\dfrac{p_v}{n_v}}\bigg)\label{keystep}.
\end{eqnarray}
The result follows by using (\ref{decompo}), (\ref{terme_A}), (\ref{terme_B}), (\ref{keystep}) and Assumption (H\ref{H4}):
$$
\frac{1}{N_v}\left| \widehat{t}_{\rm{greg}} -t_y \right|\leqslant\bigg\rvert\dfrac{1}{N_v} \sum_{i \in U_v}  \alpha_iy_i\bigg\rvert+\bigg\rvert \sum_{j=1}^{p_v} \widehat{\beta}_j b_j\bigg\rvert =\mathcal{O}_{\rm p}\left(\frac{1}{\sqrt{n}_v}\right)+\mathcal{O}_{\rm p}\left(\sqrt{\frac{p^3_v}{n_v}}\right)=\mathcal{O}_{\rm p}\left(\sqrt{\frac{p^3_v}{n_v}}\right).
$$	
%
 \subsection*{Proof of Result \ref{res2} }
 From the proof of result (\ref{res1}), we only need to show that $||\hat{\boldsymbol{\beta}}_{\rm pen}||_2=\mathcal O_{\rm p}(p_v)$ or $||\hat{\boldsymbol{\beta}}_{\rm pen}||_1=\mathcal O_{\rm p}(p_v),$ where $\hat{\boldsymbol{\beta}}_{\rm pen}$ is one of the penalized regression coefficient: ridge, lasso and elastic-net. 

\noindent Consider first the ridge regression coefficient, $\hat{\boldsymbol{\beta}}_{\rm ridge}. $ The ridge regression estimator has the advantage of having an explicit expression. We will show that $||\hat{\boldsymbol{\beta}}_{\rm ridge}||_2<||\hat{\boldsymbol{\beta}}||_2$ for $\lambda>0.$ Let denote $\hat T_{\lambda}=\boldsymbol{X}_{S_v}^\top\boldsymbol{\Pi}^{-1}_{S_v} \boldsymbol{X}_{S_v}+\lambda\boldsymbol{I}_{p_v}=\sum_{i \in S_v} \frac{\bx_i \bx_i^\top}{\pi_i}+\lambda\boldsymbol{I}_{p_v}$ sample counterpart of $T_{\lambda}=\boldsymbol{X}_{U_v}^\top \boldsymbol{X}_{U_v}+\lambda\boldsymbol{I}_{p_v}=\sum_{i \in U_v} \bx_i \bx_i^\top+\lambda\boldsymbol{I}_{p_v}. $  Moreover, let $\hat \lambda_1\geq \hat \lambda_2\geq\ldots \geq\hat \lambda_{p_v}$ be the eigenvalues of $\sum_{i \in S_v} \bx_i \bx_i^\top/\pi_i$ in decreasing order and $\hat{\mathbf{v}}_j$ the orthonormal corresponding eigenvectors, $ j=1, \ldots, p_v.$ Then, the eigenvalues of the matrix $\hat T_{\lambda}$
are $\hat \lambda_1+\lambda\geq \hat \lambda_2+\lambda\geq\ldots \geq\hat \lambda_{p_v}+\lambda\geq \lambda >0$ with the same eigenvectors $\hat{\mathbf{v}}_j, j=1, \ldots, p_v.$ Using the same arguments as those used in \cite{hoerl_kennard_1970}, we obtain $\hat{\boldsymbol{\beta}}_{\rm ridge}=\sum_{j=1}^{p_v}(\hat{\lambda}_j+\lambda)^{-1}\hat{\mathbf{v}}_j\hat{\mathbf{v}}^{\top}_j\boldsymbol{X}_{S_v}^\top\boldsymbol{\Pi}^{-1}_{S_v}\mathbf{y}_{S_v}$
and $\hat{\boldsymbol{\beta}}=\sum_{j=1}^{p_v}(\hat{\lambda}_j)^{-1}\hat{\mathbf{v}}_j\hat{\mathbf{v}}^{\top}_j\boldsymbol{X}_{S_v}^\top\boldsymbol{\Pi}^{-1}_{S_v}\mathbf{y}_{S_v}. $ Let denote by $c_j=\hat{\mathbf{v}}^{\top}_j\boldsymbol{X}_{S_v}^\top\boldsymbol{\Pi}^{-1}_{S_v}\mathbf{y}_{S_v}\in \mathbf{R},$ then
$$
||\hat{\boldsymbol{\beta}}_{\rm ridge}||_2^2=\sum_{j=1}^{p_v}\frac{c_j^2}{(\hat{\lambda}_j+\lambda)^2}<||\hat{\boldsymbol{\beta}}||_2^2=\sum_{j=1}^{p_v}\frac{c_j^2}{(\hat{\lambda}_j)^2}\quad \mbox{for}\quad \lambda>0.
$$
It follows that $||\hat{\boldsymbol{\beta}}_{\rm ridge}||_2< ||\hat{\boldsymbol{\beta}}||_2\leq ||\hat{\boldsymbol{\beta}}||_1=\mathcal{O}_{\rm p}(p_v)$ and we get  $||\hat{\boldsymbol{\beta}}_{\rm ridge}||_2=\mathcal{O}_{\rm p}(p_v). $

\noindent We now consider the lasso regression estimator,   $\hat{\boldsymbol{\beta}}_{\rm lasso},$ which minimizes the design-based version of the optimization problem given in (\ref{lasso}):
$$
\hat{\boldsymbol{\beta}}_{\rm lasso}=\argminA_{\boldsymbol{\beta} \in \mathbb{R}^p}\sum_{i\in S_v}\frac{1}{\pi_i}(y_i-\mathbf{x}_i^{\top}\boldsymbol{\beta})^2+\lambda||\boldsymbol{\beta}||_1.
$$
The lasso-estimator $\hat{\boldsymbol{\beta}}_{\rm lasso}$ may be also obtained as the solution of a constrained optimization problem:
\begin{eqnarray*}
\min_{\boldsymbol{\beta} \in \mathbb{R}^p}\sum_{i\in S_v}\frac{1}{\pi_i}(y_i-\mathbf{x}_i^{\top}\boldsymbol{\beta})^2
\end{eqnarray*}
\mbox{under the constraint}
\begin{eqnarray*}
\quad ||\boldsymbol{\beta}||_1\leq C,
\end{eqnarray*}
for some small enough constant $C>0. $ If the ordinary least-square estimator $\widehat{\boldsymbol{\beta}}$ satisfies the constraint, namely if $||\widehat{\boldsymbol{\beta}}||_1\leq C,$  then the solution of the constrained optimization problem is  $\widehat{\boldsymbol{\beta}}_{\rm lasso}=\widehat{\boldsymbol{\beta}}; $ otherwise, if $||\widehat{\boldsymbol{\beta}}||_1> C,$ then the solution  $\widehat{\boldsymbol{\beta}}_{\rm lasso}$ will be different from the least-square estimator $\widehat{\boldsymbol{\beta}}$ and $||\widehat{\boldsymbol{\beta}}_{\rm lasso}||_1\leq C<||\widehat{\boldsymbol{\beta}}||_1. $ So, in both cases, we have $||\widehat{\boldsymbol{\beta}}_{\rm lasso}||_1\leq||\widehat{\boldsymbol{\beta}}||_1=\mathcal O_{\rm p}(p_v).$

\noindent Finally, consider the elastic-net regression estimator,   $\hat{\boldsymbol{\beta}}_{\rm en}. $ Consider the following objective functions:
 \begin{align*}
 	\mathcal{L}_{ols} (\boldsymbol{\beta}) &= \sum_{i\in S_v}\frac{1}{\pi_i}(y_i-\mathbf{x}_i^{\top}\boldsymbol{\beta})^2\\
 	 	 	\mathcal{L}_{en} (\boldsymbol{\beta}) &= \sum_{i\in S_v}\frac{1}{\pi_i}(y_i-\mathbf{x}_i^{\top}\boldsymbol{\beta})^2 + \lambda_1 \rvert\rvert \boldsymbol{\beta} \rvert\rvert_1 + \lambda_2 \rvert\rvert \boldsymbol{\beta} \rvert\rvert_2^2 =  	\mathcal{L}_{ols} (\boldsymbol{\beta}) + \lambda_1 \rvert\rvert \boldsymbol{\beta} \rvert\rvert_1 + \lambda_2 \rvert\rvert \boldsymbol{\beta} \rvert\rvert_2^2,
 \end{align*}
 where $\lambda_1=\lambda \alpha$ and $\lambda_2=\lambda(1-\alpha)$ with $\lambda>0$ and $\alpha\in (0,1). $ The cases $\alpha=0$ and $\alpha=1$ lead, respectively,  to the ridge and lasso regression estimators which have been discussed above.
The ordinary least squares estimator $\widehat{\boldsymbol{\beta}}$ verifies $\widehat{\boldsymbol{\beta}} =\argminA_{\boldsymbol{\beta} \in \mathbb{R}^p} 	\mathcal{L}_{ols} (\boldsymbol{\beta}) $ and the elastic-net estimator verifies $ \widehat{\boldsymbol{\beta}}_{en}=   \argminA_{\boldsymbol{\beta} \in \mathbb{R}^p}	\mathcal{L}_{en} (\boldsymbol{\beta}) $.  
 Since $\widehat{\boldsymbol{\beta}}$ minimizes  $ \mathcal{L}_{ols}(\boldsymbol{\beta})$, we have $\mathcal{L}_{ols} (\widehat{\boldsymbol{\beta}})  \leqslant 	\mathcal{L}_{ols} (\widehat{\boldsymbol{\beta}}_{en}) $. Similarly, we have $\mathcal{L}_{en} (\widehat{\boldsymbol{\beta}}_{en})  \leqslant 	\mathcal{L}_{en} (\widehat{\boldsymbol{\beta}}_{ols})$. Therefore, the following inequalities hold:
\begin{eqnarray*}
	\mathcal{L}_{ols} (\widehat{\boldsymbol{\beta}}) + \lambda_1 \rvert\rvert \widehat{\boldsymbol{\beta}}_{en} \rvert\rvert_1 + \lambda_2 \rvert\rvert \widehat{\boldsymbol{\beta}}_{en}  \rvert\rvert_2^2
&\leqslant& 	\mathcal{L}_{ols} (\widehat{\boldsymbol{\beta}}_{en}) + \lambda_1 \rvert\rvert \widehat{\boldsymbol{\beta}}_{en}  \rvert\rvert_1 + \lambda_2 \rvert\rvert \widehat{\boldsymbol{\beta}}_{en}  \rvert\rvert_2^2 =\mathcal{L}_{en} (\widehat{\boldsymbol{\beta}}_{en})
 \nonumber \\
 &\leqslant & \mathcal{L}_{ols} (\widehat{\boldsymbol{\beta}}) + \lambda_1 \rvert\rvert\widehat{\boldsymbol{\beta}}\rvert\rvert_1 + \lambda_2 \rvert\rvert\widehat{\boldsymbol{\beta}} \rvert\rvert_2^2=	\mathcal{L}_{en} (\widehat{\boldsymbol{\beta}}_{ols}), \label{eq2}
\end{eqnarray*}
which implies
\begin{align}
\lambda_1 \rvert\rvert \widehat{\boldsymbol{\beta}}_{en} \rvert\rvert_1 + \lambda_2 \rvert\rvert \widehat{\boldsymbol{\beta}}_{en} \rvert\rvert_2^2 &\leqslant \lambda_1 \rvert\rvert\widehat{\boldsymbol{\beta}} \rvert\rvert_1 + \lambda_2 \rvert\rvert\widehat{\boldsymbol{\beta}} \rvert\rvert_2^2. \label{eq3}
\end{align}
Furthermore, since $\lambda_1 > 0$, we can write
\begin{align} \label{eq4}
	\lambda_2 \rvert\rvert \widehat{\boldsymbol{\beta}}_{en}  \rvert\rvert_2^2 &\leqslant  \lambda_1 \rvert\rvert \widehat{\boldsymbol{\beta}}_{en} \rvert\rvert_1 + \lambda_2 \rvert\rvert \widehat{\boldsymbol{\beta}}_{en} \rvert\rvert_2^2.
\end{align}
Using  (\ref{eq3}), (\ref{eq4}) and the fact that $\rvert\rvert\widehat{\boldsymbol{\beta}} \rvert\rvert_2 \leqslant\rvert\rvert\widehat{\boldsymbol{\beta}} \rvert\rvert_1,$ we obtain
\begin{align*}
\lambda_2 \rvert\rvert \widehat{\boldsymbol{\beta}}_{en}  \rvert\rvert_2^2 &\leqslant \lambda_1 \rvert\rvert\widehat{\boldsymbol{\beta}}\rvert\rvert_1 + \lambda_2 \rvert\rvert\widehat{\boldsymbol{\beta}}\rvert\rvert_2^2 \leqslant  \lambda_1 \rvert\rvert\widehat{\boldsymbol{\beta}} \rvert\rvert_1 + \lambda_2 \rvert\rvert\widehat{\boldsymbol{\beta}} \rvert\rvert_1^2
\end{align*}
which implies
\begin{align*}
\rvert\rvert \widehat{\boldsymbol{\beta}}_{en} \rvert\rvert_2^2 \leqslant  \dfrac{\alpha}{1-\alpha} \rvert\rvert\widehat{\boldsymbol{\beta}} \rvert\rvert_1 +  \rvert\rvert\widehat{\boldsymbol{\beta}}\rvert\rvert_1^2 = \mathcal{O}_{\rm{p}}(p_v^2)
\end{align*}
and so, $\rvert\rvert \widehat{\boldsymbol{\beta}}_{en} \rvert\rvert_2= \mathcal{O}_{\rm{p}}(p_v).$

 \subsection*{Proof of Result \ref{result_consist_ridge} }
\begin{enumerate}
\item 

As in the proof of result (\ref{res2}), we consider the eigenvalues of the matrix $\hat T_{\lambda}$
in decreasing order: $\hat \lambda_1+\lambda\geq \hat \lambda_2+\lambda\geq\ldots \geq\hat \lambda_{p_v}+\lambda\geq \lambda >0$. The matrix  $\hat T_{\lambda}$ is always invertible and the eigenvalues of $\hat T_{\lambda}^{-1}$ are $0<(\hat \lambda_1+\lambda)^{-1}\leq (\hat \lambda_2+\lambda)^{-1}\leq\ldots \leq(\hat \lambda_{p_v}+\lambda)^{-1}\leq \lambda^{-1}.$ We then obtain 
\begin{eqnarray}
\rvert\rvert\hat T_{\lambda}^{-1}\rvert\rvert_2\leqslant \lambda^{-1},\label{borne_inv_matrix}
\end{eqnarray}
where $||\cdot||_2$ is the spectral norm matrix defined for a squared $p\times p$ matrix $\mathbf A$ as $||\mathbf{A}||_2=\sup_{\mathbf{x}\in \mathbf{R}^p, ||\mathbf{x}||_2\neq 0}||\mathbf{A}\mathbf{x}||_2/||\mathbf{x}||_2. $ For a symmetric and positive definite matrix $\mathbf{A}$, we have that $||\mathbf{A}||_2=\lambda_{max}(\mathbf{A})$ where $\lambda_{max}(\mathbf{A})$ is the largest eigenvalue of $\mathbf{A}.$ Now, we can write
\begin{eqnarray}
\frac{1}{N^2_v}\bigg\rvert\bigg\rvert\sum_{i\in S_v}\frac{\bx_i y_i}{\pi_i}\bigg\rvert\bigg\rvert_2^2&=&\frac{1}{N^2_v}\sum_{i\in U_v}\sum_{\ell\in U_v}\bx^{\top}_i \bx_{\ell}\frac{y_iI_i}{\pi_i}\frac{y_{\ell}I_{\ell}}{\pi_{\ell}}=\frac{1}{N^2_v}\mathcal Y^{\top}\boldsymbol{X}_{U_v}\boldsymbol{X}^{\top}_{U_v}\mathcal Y\nonumber\\
&\leqslant&\frac{1}{N_v}\rvert\rvert\mathcal Y\rvert\rvert_2^2\frac{1}{N_v}\rvert\rvert\boldsymbol{X}_{U_v}\boldsymbol{X}^{\top}_{U_v}\rvert\rvert_2,\nonumber
\end{eqnarray}
where $\mathcal Y^{\top}=\displaystyle\left(\frac{y_iI_i}{\pi_i}\right)_{i\in U_v}.$ The symmetric and positive semi-definite  $N_v\times N_v$ matrix $\boldsymbol{X}_{U_v}\boldsymbol{X}^{\top}_{U_v}$ has the same non-null eigenvalues as those of the  positive definite $p_v\times p_v$ matrix $\boldsymbol{X}^{\top}_{U_v}\boldsymbol{X}_{U_v}, $. Therefore, 
$$
\frac{1}{N_v}\rvert\rvert\boldsymbol{X}_{U_v}\boldsymbol{X}^{\top}_{U_v}\rvert\rvert_2=\frac{1}{N_v}\lambda_{max}(\boldsymbol{X}^{\top}_{U_v}\boldsymbol{X}_{U_v})\leqslant \tilde C.
$$
Using Assumptions (H\ref{H1}) and (H\ref{H3}), we have
\begin{eqnarray*}
\frac{1}{N^2_v}\bigg\rvert\bigg\rvert\sum_{i\in S_v}\frac{\bx_i y_i}{\pi_i}\bigg\rvert\bigg\rvert_2^2\leqslant \frac{\tilde C}{N_v}\rvert\rvert\mathcal Y\rvert\rvert_2^2=\frac{\tilde C}{N_v}\sum_{i\in U_v}\frac{y^2_iI_i}{\pi^2_i}\leqslant \frac{\tilde C}{c^2N_v}\sum_{i\in U_v}y^2_i=\mathcal{O}(1).
\end{eqnarray*}
 Finally, using also the fact that $N_v/\lambda=O(1),$ we have
$$
\rvert \rvert \widehat{\boldsymbol{\beta}}_{\rm{ridge}} \rvert \rvert _2^2\leqslant\rvert\rvert\hat T_{\lambda}^{-1}\rvert\rvert^2_2\bigg\rvert\bigg\rvert\sum_{i\in S_v}\frac{\bx_i y_i}{\pi_i}\bigg\rvert\bigg\rvert_2^2\leqslant N^2_v\lambda^{-2}\bigg\rvert\bigg\rvert\frac{1}{N^2_v}\sum_{i\in S_v}\frac{\bx_i y_i}{\pi_i}\bigg\rvert\bigg\rvert_2^2=\mathcal{O}(1).
$$
It follows that
\begin{eqnarray}
\mathbb{E}_{\rm p}\left[\rvert \rvert \widehat{\boldsymbol{\beta}}_{\rm{ridge}}  \rvert \rvert _2^2 \right] =\mathcal O(1).\label{borne_beta_ridge}
\end{eqnarray}

To obtain the $L^1$ design-consistency of the ridge model-assisted estimator, we write as in the proof of Result \ref{res1}:
\begin{eqnarray*}
\dfrac{1}{N_v}	\left( \widehat{t}_{\rm{ridge}} -t_y \right) &=& \dfrac{1}{N_v} \sum_{i \in U_v}  \alpha_iy_i - \sum_{j=1}^{p_v}  b_j\widehat{\beta}_{j,\rm{ridge}}\\
&=&  \dfrac{1}{N_v} \sum_{i \in U_v}  \alpha_iy_i-\frac{1}{N_v}\left(\sum_{i \in U_v}  \alpha_i\bx_i\right)^{\top}\widehat{\boldsymbol{\beta}}_{\rm{ridge}}
\end{eqnarray*}
and
\begin{eqnarray*}
\mathbb{E}_{\rm p}\bigg|\dfrac{1}{N_v}\left( \widehat{t}_{\rm{ridge}} -t_y \right)\bigg|&\leqslant&\mathbb{E}_{\rm p}\bigg|\dfrac{1}{N_v} \sum_{i \in U_v}  \alpha_iy_i\bigg|+\sqrt{\mathbb{E}_{\rm p}\left(\frac{1}{N^2_v}\bigg|\bigg|\sum_{i \in U_v}  \alpha_i\bx_i\bigg|\bigg|_2^2\right)\mathbb{E}_{\rm p}||\widehat{\boldsymbol{\beta}}_{\rm{ridge}} ||^2_2}\nonumber\\
&=& \mathcal{O}\left(\sqrt{\frac{1}{n_v}}\right)+\mathcal{O}\left(\sqrt{\frac{p_v}{n_v}}\right)=\mathcal{O}\left(\sqrt{\frac{p_v}{n_v}}\right)
\end{eqnarray*}
by (\ref{borne_a_v2}), (\ref{borne_x}), (\ref{borne_beta_ridge}).
\item We can write
\begin{eqnarray}
\widehat{\boldsymbol{\beta}}_{\rm{ridge}} -\widetilde{\boldsymbol{\beta}}_{\rm{ridge}}=\widehat T^{-1}_{\lambda}\left(\sum_{i\in S_v}\frac{E_{i\lambda}}{\pi_i}-\sum_{i\in U_v}E_{i\lambda}\right),\label{decomp_beta_ridge}
\end{eqnarray}
where $E_{i\lambda}=\bx_i(y_i-\bx^{\top}_i\tilde{\boldsymbol{\beta}}_{\rm{ridge}})$ with $\sum_{i\in U_v}E_{i\lambda}=\lambda\boldsymbol{I}_{p_v}.$ Using the same arguments as those used in the proof of Result \ref{res1}, we get
\begin{eqnarray}
\frac{1}{N_v^2}\mathbb{E}_{\rm{p}}\bigg\rvert\bigg\rvert\sum_{i\in S_v}\frac{E_{i\lambda}}{\pi_i}-\sum_{i\in U_v}E_{i\lambda}\bigg\rvert\bigg\rvert_2^2\leqslant \left(\frac{1}{cN_v}+\frac{n_v\max_{i,\ell\in U_v,i\neq \ell}|\pi_{i\ell}-\pi_i\pi_{\ell}|}{c^2 n_v}\right)\frac{1}{N_v}\sum_{i\in U_v}||E_{i\lambda}||_2^2.\nonumber\\
\label{ordre_E_1}
\end{eqnarray}
Furthermore,
\begin{eqnarray}
\frac{1}{N_v}\sum_{i\in U_v}||E_{i\lambda}||_2^2\leqslant \frac{2C_2p_v}{N_v}\left(\sum_{i\in U_v}y^2_i+\sum_{i\in U_v}(\bx^{\top}_i\tilde{\boldsymbol{\beta}}_{\rm{ridge}})^2\right)=\mathcal O(p_v)\label{ordre_E_2}
\end{eqnarray}
by Assumptions (H\ref{H1}) and (H\ref{Hnew}) and the fact that
\begin{eqnarray}
\frac{1}{N_v}\sum_{i\in U_v}(\bx^{\top}_i\tilde{\boldsymbol{\beta}}_{\rm{ridge}})^2=\tilde{\boldsymbol{\beta}}^{\top}_{\rm{ridge}}\left(\frac{1}{N_v}\sum_{i\in U_v}\bx_i\bx^{\top}_i\right)\tilde{\boldsymbol{\beta}}_{\rm{ridge}}\leqslant ||\tilde{\boldsymbol{\beta}}_{\rm{ridge}}||_2^2\frac{1}{N_v}\rvert\rvert\boldsymbol{X}^{\top}_{U_v}\boldsymbol{X}_{U_v}\rvert\rvert_2=\mathcal O(1).\nonumber\\
\label{sum_predict_pop}
\end{eqnarray}
To obtain the above inequality, we have also used the fact that $ ||\tilde{\boldsymbol{\beta}}_{\rm{ridge}}||_2=\mathcal O(1)$ which can be proved by using the same arguments as the ones used for showing that $ ||\widehat{\boldsymbol{\beta}}_{\rm{ridge}}||_2=\mathcal O(1)$ in point (1). Expressions (\ref{ordre_E_1}) and (\ref{ordre_E_2}) lead to
\begin{eqnarray}
\frac{1}{N_v^2}\mathbb{E}_{\rm{p}}\bigg\rvert\bigg\rvert\sum_{i\in S_v}\frac{E_{i\lambda}}{\pi_i}-\sum_{i\in U_v}E_{i\lambda}\bigg\rvert\bigg\rvert_2^2=\mathcal O\left(\frac{p_v}{n_v}\right).\label{conv_E_ilambda}
\end{eqnarray}
The result follows from (\ref{decomp_beta_ridge}), (\ref{conv_E_ilambda}) and the fact that $||N_v\hat T_{\lambda}^{-1}||_2=\mathcal O(1):$
\begin{eqnarray}
\mathbb{E}_{\rm{p}}\rvert\rvert\widehat{\boldsymbol{\beta}}_{\rm{ridge}} -\tilde{\boldsymbol{\beta}}_{\rm{ridge}}\rvert\rvert_2^2=\mathcal O\left(\frac{p_v}{n_v}\right).\label{L1_conv_beta_ridge}
\end{eqnarray}
\item We use the following decomposition:
\begin{eqnarray*}
\frac{1}{N_v}\left(\widehat{t}_{\rm{ridge}} -t_y\right)=\frac{1}{N_v}\left(\widehat{t}_{\rm{diff,\lambda}} -t_y\right)-\frac{1}{N_v}\left(\sum_{i\in S_v}\frac{\bx_i}{\pi_i}-\sum_{i\in U_v}\bx_i\right)^{\top}\left(\widehat{\boldsymbol{\beta}}_{\rm{ridge}} -\tilde{\boldsymbol{\beta}}_{\rm{ridge}}\right),
\end{eqnarray*}
and
\begin{eqnarray*}
\frac{1}{N_v}\left(\widehat{t}_{\rm{diff,\lambda}} -t_y\right)&=&\frac{1}{N_v}\left(\sum_{i\in S_v}\frac{y_i}{\pi_i}-\sum_{i\in U_v}y_i\right)-\frac{1}{N_v}\left(\sum_{i\in S_v}\frac{\bx_i}{\pi_i}-\sum_{i\in U_v}\bx_i\right)^{\top}\tilde{\boldsymbol{\beta}}_{\rm{ridge}}\\
&= &\frac{1}{N_v}\sum_{i\in U_v}\alpha_iy_i-\frac{1}{N_v}\sum_{i\in U_v}\alpha_i\bx_i^{\top}\tilde{\boldsymbol{\beta}}_{\rm{ridge}},
\end{eqnarray*}
where $\alpha_i=I_i/\pi_i-1, i\in U_v.$ From (\ref{borne_a_v2}), we have that $N_v^{-2}\mathbb{E}_{\rm{p}}(\sum_{i\in U_v}\alpha_iy_i)^2=\mathcal O(n_v^{-1})$ and we can get $N^{-2}_v\mathbb{E}_{\rm{p}}\left(\sum_{i\in U_v}\alpha_i\bx^{\top}_i\tilde{\boldsymbol{\beta}}_{\rm{ridge}}\right)^2=\mathcal O(n_v^{-1})$ by using similar arguments as those used in the proof of Result \ref{res1} and (\ref{sum_predict_pop}).  We obtain
$$
\frac{1}{N^2_v}\mathbb{E}_{\rm{p}}\left(\widehat{t}_{\rm{diff,\lambda}} -t_y\right)^2=\mathcal O\left(\frac{1}{n_v}\right).
$$
The results follows  since
\begin{eqnarray*}
\frac{1}{N_v}\mathbb{E}_p \bigg\rvert \widehat{t}_{\rm ridge} -t_y\bigg\rvert &\leqslant& \frac{1}{N_v}\mathbb{E}_p \bigg\rvert \widehat{t}_{\rm diff,\lambda} -t_y\bigg\rvert+\sqrt{\frac{1}{N^2_v}\mathbb{E}_p\bigg\rvert \bigg\rvert\sum_{i\in S_v}\frac{\bx_i}{\pi_i}-\sum_{i\in U_v}\bx_i\bigg\rvert\bigg\rvert_2^2\mathbb{E}_p\bigg\rvert\bigg\rvert\widehat{\boldsymbol{\beta}}_{\rm{ridge}} -\tilde{\boldsymbol{\beta}}_{\rm{ridge}}\bigg\rvert\bigg\rvert_2^2}\\
&=&\mathcal O\left(\frac{1}{\sqrt{n_v}}\right)+\mathcal O\left(\frac{p_v}{n_v}\right)
\end{eqnarray*}
by using (\ref{borne_x}) and (\ref{L1_conv_beta_ridge}).
\end{enumerate}

 \subsection*{Proof of Proposition \ref{lasso_orthogon}}
From the proof of Result \ref{res1} (more specifically, Equations \ref{terme_B} and \ref{borne_x}), we need to show that $\sum_{i\in U_v}||\mathbf{x}_i||^2_2/N_v=\mathcal O(p_v)$ and that $||\widehat{\boldsymbol{\beta}}||_2=\mathcal O_{\rm{p}}(1)$. The same result holds for $\widehat{\boldsymbol{\beta}}_{\rm{lasso}}$ and $\widehat{\boldsymbol{\beta}}_{\rm{en}}.$ We have $\sum_{i\in U_v}||\mathbf{x}_i||^2_2/N_v=\sum_{j=1}^{p_v}\sum_{i\in U_v}x_{ij}^2/N_v\leq p_v\sqrt{C_3}=\mathcal O(p_v)$ under the assumption of uniformly bounded forth moment of $X_j, j=1, \ldots, p_v. $

We first show that, under the assumed orthogonality condition, $||\widehat{\boldsymbol{\beta}}_{\rm{lasso}}||_2\leq ||\widehat{\boldsymbol{\beta}}||_2$, $||\widehat{\boldsymbol{\beta}}_{\rm{en}}||_2\leq ||\widehat{\boldsymbol{\beta}}||_2$ and also $||\widehat{\boldsymbol{\beta}}||_2=\mathcal O_{\rm{p}}(1).$\\
 Consider again the objective function $\mathcal{L}_{ols} (\boldsymbol{\beta})$ as in the proof of Result \ref{res2}. We can write
 \begin{eqnarray}
 \mathcal{L}_{ols} (\boldsymbol{\beta}) &= & \sum_{i\in S_v}\frac{1}{\pi_i}(y_i-\mathbf{x}_i^{\top}\boldsymbol{\beta})^2= \sum_{i\in S_v}(\tilde{y}_i-\tilde{\bx}^{\top}_i\boldsymbol{\beta})^2
 \end{eqnarray}
 where $\tilde{y}_i=y_i/\sqrt{\pi_i}$ and $\tilde{\bx}_i=(\tilde x_{ij})_{j=1}^{p_v}=\bx_i/\sqrt{\pi_i}$ for all $i\in S_v.$  Let $\tilde{\mathbf{X}}_{S_v}=\boldsymbol{\Pi}^{-1/2}_{S_v}\mathbf{X}_{S_v}=(\tilde{\bx}^{\top}_i)_{i\in S_v}=(\tilde{\mathbf{X}}_1, \ldots,\tilde{\mathbf{X}}_{p_v}). $ The  columns of $\tilde{\mathbf{X}}_{S_v},$ denoted by $\tilde{\mathbf{X}}_j, j=1, \ldots, p_v$ are assumed to be orthogonal. This means that $\tilde{\mathbf{X}}^{\top}_j\tilde{\mathbf{X}}_k=0$ for $j\neq k. $ The ordinary least-square estimator  $\widehat{\boldsymbol{\beta}}$ is given by
 $$
 \widehat{\boldsymbol{\beta}}=(\tilde{\mathbf{X}}^{\top}_{S_v}\tilde{\mathbf{X}}_{S_v})^{-1}\tilde{\mathbf{X}}_{S_v}^{\top}\tilde{\mathbf y}_{S_v}.
 $$
Under the  orthogonality condition, $\tilde{\mathbf{X}}^{\top}_{S_v}\tilde{\mathbf{X}}_{S_v}$ is a diagonal matrix with diagonal elements given by $||\tilde{\mathbf{X}}_j||_2^2=\sum_{i\in S_v}\tilde{x}^2_{ij}=\sum_{i\in S_v}\frac{x^2_{ij}}{\pi_i}, $ which corresponds to the Horvitz-Thompson estimator of $\sum_{i\in U_v}x^2_{ij}. $ Therefore,  $\widehat{\boldsymbol{\beta}}=(\hat{\beta}_{j})_{j\in S_v}$ and the $j$-th coordinate is given by $\hat{\beta}_{j}=(\sum_{i\in S_v}\tilde{x}^2_{ij})^{-1}\sum_{i\in S_v}\tilde{x}_{ij}\tilde{y}_i. $

 The lasso estimator $\hat{\boldsymbol{\beta}}_{\rm{lasso}}=(\hat{\beta}_{j,\rm{lasso}})_{j=1}^{p_v}$ as well as the elastic-net estimator $\hat{\boldsymbol{\beta}}_{\rm{en}}=(\hat{\beta}_{j,\rm{en}})_{j=1}^{p_v}$ are obtained by using the cyclic soft-thresholding algorithm  \citep{hastie_tibshirani_friedman_2011}:
 $$
 \hat{\beta}_{j,\rm{lasso}}=\frac{\mathcal{S}_{\lambda}(\sum_{i\in S_v}r_{ij}\tilde{x}_{ij})}{\sum_{i\in S_v}\tilde{x}^2_{ij}}
 $$
 and
 $$
 \hat{\beta}_{j,\rm{en}}=\frac{\mathcal{S}_{\lambda\alpha}(\sum_{i\in S_v}r_{ij}\tilde{x}_{ij})}{\sum_{i=1}^{n_v}\tilde{x}^2_{ij}+\lambda(1-\alpha)},
 $$
 where $r_{ij}=\tilde{y}_i-\sum_{k\neq j}\tilde{x}_{ik}\hat{\beta}_k$ and $\mathcal{S}_{\lambda}(z)=sign(z)(|z|-\lambda)_+$ is the soft-thresholding function with $(|z|-\lambda)_+=|z|-\lambda$ if $|z|\geq \lambda,$ and zero otherwise. If the columns of $\tilde{\mathbf{X}}_{S_v}$ are orthogonal, then $\sum_{i\in S_v}r_{ij}\tilde{x}_{ij}=\sum_{i\in S_v}\tilde{x}_{ij}\tilde{y}_i$ and $\hat{\beta}_{j,\rm{lasso}}$ is the soft-threshold estimator of the least-square estimator $\widehat{\beta}_j$:
 $$
 \hat{\beta}_{j,\rm{lasso}}=\frac{\mathcal{S}_{\lambda}(\sum_{i\in S_v}\tilde{x}_{ij}\tilde{y}_i)}{\sum_{i\in S_v}\tilde{x}^2_{ij}}.
 $$
 The elastic-net estimator is given by
 $$
  \hat{\beta}_{j,\rm{en}}=\frac{\mathcal{S}_{\lambda\alpha}(\sum_{i\in S_v}\tilde{x}_{ij}\tilde{y}_i)}{\sum_{i\in S_v}\tilde{x}^2_{ij}+\lambda(1-\alpha)}.
 $$
It follows that
 $$
 |\hat{\beta}_{j,\rm{lasso}}|=\frac{|(|\sum_{i\in S_v}\tilde{x}_{ij}\tilde{y}_i|-\lambda)_+|}{\sum_{i\in S_v}\tilde{x}^2_{ij}}\leq \frac{|\sum_{i\in S_v}\tilde{x}_{ij}\tilde{y}_i|}{\sum_{i\in S_v}\tilde{x}^2_{ij}}=|\hat{\beta}_{j}|, \quad j=1, \ldots, p_v
 $$
and $||\widehat{\boldsymbol{\beta}}_{\rm{lasso}}||_2\leq ||\widehat{\boldsymbol{\beta}}||_2. $ Similarly, $||\widehat{\boldsymbol{\beta}}_{\rm{en}}||_2\leq ||\widehat{\boldsymbol{\beta}}||_2. $

We now show that $||\widehat{\boldsymbol{\beta}}||_2=\mathcal O_{\rm{p}}(1). $ We have
$$
||\widehat{\boldsymbol{\beta}}||_2\leq ||N_v(\tilde{\mathbf{X}}^{\top}_{S_v}\tilde{\mathbf{X}}_{S_v})^{-1}||_2\left|\left|\frac{1}{N_v}\tilde{\mathbf{X}}^{\top}_{S_v}\tilde{\mathbf y}_{S_v}\right|\right|_2.
$$
The  matrix $\tilde{\mathbf{X}}^{\top}_{S_v}\tilde{\mathbf{X}}_{S_v}$ is diagonal with diagonal elements equal to $\sum_{i\in S_v}\frac{x^2_{ij}}{\pi_i}.$ Then,
$$
||N_v(\tilde{\mathbf{X}}^{\top}_{S_v}\tilde{\mathbf{X}}_{S_v})^{-1}||_2=\max_{j=1, \ldots, p_v}\left(\frac{1}{N^{-1}_v\sum_{i\in S_v}\frac{x^2_{ij}}{\pi_i}}\right)
$$
and for all $j=1, \ldots, p_v:$
\begin{eqnarray*}
\frac{1}{N^{-1}_v\sum_{i\in S_v}\frac{x^2_{ij}}{\pi_i}}=\frac{1}{N_v^{-1}\sum_{i\in U_v}x^2_{ij}}+\mathcal O_{\rm{p}}\left(\frac{1}{\sqrt{n_v}}\right)=\mathcal O_{\rm p}(1)
\end{eqnarray*}
by using (H\ref{H2}), (H\ref{H3}) and the assumption of uniformly bounded fourth moment of $X_j, j=1,\ldots, p_v$. We have also used the fact that $1/(N_v^{-1}\sum_{i\in U_v}x^2_{ij})\leq 1/(\min_{j=1, \ldots, p_v}N_v^{-1}\sum_{i\in U_v}x^2_{ij})\leq 1/C_4=\mathcal O(1)$ for all $j=1, \ldots, p_v.$  Then,
\begin{eqnarray}
||N_v(\tilde{\mathbf{X}}^{\top}_{S_v}\tilde{\mathbf{X}}_{S_v})^{-1}||_2=\mathcal O_{\rm{p}}(1). \label{borne_XtX}
\end{eqnarray}
Now,
$$
\left|\left|\frac{1}{N_v}\tilde{\mathbf{X}}^{\top}_{S_v}\tilde{\mathbf y}_{S_v}\right|\right|^2_2\leq \frac{1}{N_v}\left|\left|\tilde{\mathbf y}_{S_v}\right|\right|_2^2 \left|\left|\frac{1}{N_v}\tilde{\mathbf{X}}_{S_v}\tilde{\mathbf{X}}^{\top}_{S_v}\right|\right|_2.
$$
We have
$$
\left|\left|\frac{1}{N_v}\tilde{\mathbf{X}}_{S_v}\tilde{\mathbf{X}}^{\top}_{S_v}\right|\right|_2=\left|\left|\frac{1}{N_v}\tilde{\mathbf{X}}^{\top}_{S_v}\tilde{\mathbf{X}}_{S_v}\right|\right|_2=\max_{j=1, \ldots, p_v}\left(\frac{1}{N_v}\sum_{i\in S_v}\frac{x^2_{ij}}{\pi_i}\right)\leq \max_{j=1, \ldots, p_v}\left(\frac{1}{N_v}\sum_{i\in U_v}x^2_{ij}\right)\leq \sqrt{C_3}
$$
and
\begin{eqnarray*}
\frac{1}{N_v}||\tilde{\mathbf y}_{S_v}||_2^2=\frac{1}{N_v}\sum_{i\in S_v}\frac{y^2_i}{\pi^2_i}\leq \frac{1}{c^2N_v}\sum_{i\in U_v}y^2_i\leq \frac{C_1}{c^2}
\end{eqnarray*}
by Assumption (H\ref{H1}). So, $||\frac{1}{N_v}\tilde{\mathbf{X}}_{S_v}\tilde{\mathbf y}_{S_v}||_2=\mathcal O(1)$ and combined with (\ref{borne_XtX}), we obtain  $||\widehat{\boldsymbol{\beta}}||_2=\mathcal O_{\rm{p}}(1). $

 \subsection*{Proof of Result \ref{res3} }
	Again, we use the decomposition (\ref{decompo}) as in the proof of Result \ref{res1}:
\begin{eqnarray*}
\frac{1}{N_v}(\widehat{t}_{\rm{tree}} -t_y)=\dfrac{1}{N_v} \sum_{i \in U_v}  \alpha_i y_i-\sum_{j=1}^{T_v}b_j\hat{\beta}_j,
\end{eqnarray*}	
 where $b_j=\dfrac{1}{N_v} \sum_{i \in U_v} \alpha_i z_{ij}$ with $z_{ij}=\mathds{1}\left(\bx_i \in A_j\right)$ for $j=1, 2, ..., T_v.$ We have
\begin{eqnarray}
\mathbb{E}_{\rm{p}} \bigg[ \dfrac{1}{N_v}\bigg\rvert \widehat{t}_{\rm{tree}} -t_y\bigg\rvert  \bigg]
\leqslant \mathbb{E}_{\rm{p}} \bigg[ \dfrac{1}{N_v}\bigg\rvert\sum_{i \in U_v}  \alpha_i y_i\bigg\rvert  \bigg]  +\mathbb{E}_{\rm{p}} \bigg\rvert\sum_{j=1}^{T_v}b_j\hat{\beta}_j\bigg\rvert.\label{decomp_tree}
\end{eqnarray}
We obtain from the proof of Result \ref{res1}, (see Equation (\ref{borne_a_v2})) that
\begin{eqnarray}
\mathbb{E}_{\rm{p}} \bigg[ \dfrac{1}{N_v}\bigg\rvert\sum_{i \in U_v}  \alpha_i y_i\bigg\rvert  \bigg] =\mathcal O\left(\sqrt{\frac{1}{n_v}}\right). \label{terme_A_tree}
\end{eqnarray}
Now, consider the second term on the right-side of (\ref{decomp_tree}). We have
\begin{eqnarray}
\mathbb{E}_{\rm{p}}  \bigg\rvert \sum_{j=1}^{p_v} \widehat{\beta}_j b_j\bigg\rvert \leqslant\sqrt{\mathbb{E}_{\rm{p}} \left(\sum_{j=1}^{p_v} \widehat{\beta}_j^2\right)\mathbb{E}_{\rm{p}} \left(\sum_{j=1}^{p_v}  b_j^2\right)}=\sqrt{\left(\mathbb{E}_{\rm{p}}  ||\widehat{\boldsymbol{\beta}}||^2_2\right)\left[\mathbb{E}_{\rm{p}} \left( \sum_{j=1}^{p_v}  b_j^2\right)\right]}.\label{terme_B_tree}
\end{eqnarray}
We show first that
\begin{eqnarray}
\mathbb{E}_{\rm{p}}  ||\widehat{\boldsymbol{\beta}}||^2_2=\mathcal{O}\left(\frac{n_{v}}{n_{0v}}\right).\label{borne_beta_tree}
 \end{eqnarray}
We have
$$
 ||\widehat{\boldsymbol{\beta}}||^2_2\leq ||(\boldsymbol{Z}^{\top}_{S_v}\boldsymbol{\Pi}_{S_v}^{-1}\boldsymbol{Z}_{S_v})^{-1}||^2_2||\boldsymbol{Z}^{\top}_{S_v}\boldsymbol{\Pi}_{S_v}^{-1}\boldsymbol{y}_{S_v}||_2^2
$$
and $||(\boldsymbol{Z}^{\top}_{S_v}\boldsymbol{\Pi}_{S_v}^{-1}\boldsymbol{Z}_{S_v})^{-1}||_2=\mathcal{O}(n^{-1}_{0v})$ since the matrix   $\boldsymbol{Z}^{\top}_{S_v}\boldsymbol{\Pi}_{S_v}^{-1}\boldsymbol{Z}_{S_v}$ is diagonal  with diagonal elements equal to $\sum_{i\in S_v}\frac{1}{\pi_i}\mathds{1} \left(\bx_i \in A_j\right)\geq n_0>0$ for all $j=1, \ldots T_v$ by the stopping criterion and not depending on the sampling design. Moreover, we get following the same lines as  in the proofs of result (\ref{result_consist_ridge}) or proposition (\ref{lasso_orthogon}):
\begin{eqnarray*}
\frac{1}{N_v}||\boldsymbol{Z}^{\top}_{S_v}\boldsymbol{\Pi}_{S_v}^{-1}\boldsymbol{y}_{S_v}||_2^2\leq \frac{1}{N_v}||\boldsymbol{y}^{\top}_{S_v}\boldsymbol{\Pi}_{S_v}^{-1}||^2_2||\boldsymbol{Z}_{S_v}\boldsymbol{Z}^{\top}_{S_v}||_2\leq\frac{1}{c^2N_v}\left(\sum_{i\in U_v}y^2_i\right)||\boldsymbol{Z}^{\top}_{S_v}\boldsymbol{Z}_{S_v}||_2=\mathcal{O}(n_{0v}),
\end{eqnarray*}
using Assumption (H\ref{H1}) and the fact that the matrix $\boldsymbol{Z}^{\top}_{S_v}\boldsymbol{Z}_{S_v}$ is diagonal with diagonal elements equal to  $\sum_{i\in S_v}\mathds{1} \left(\bx_i \in A_j\right)\leqslant 2n_{0v}-1$  for all $j=1, \ldots T_v$ by the stopping criterion. therefore,
 $||\boldsymbol{Z}^{\top}_{S_v}\boldsymbol{Z}_{S_v}||_2=\max_{j=1, \ldots, T_v}(\sum_{i\in S_v}\mathds{1} \left(\bx_i \in A_j\right))= \mathcal O(n_{0v}). $ Finally, we obtain $||\widehat{\boldsymbol{\beta}}||^2_2 \leqslant\mathcal{O}(n^{-2}_{0v})\mathcal{O}(N_vn_{0v})=\mathcal{O}(n_{v}/n_{0v}).$

Now,
 \begin{eqnarray}
	\mathbb{E}_{\rm{p}} \left( \sum_{j=1}^{p_v}  b_j^2\right)&=&\dfrac{1}{N_v^2}\sum_{j=1}^{T_v} \sum_{i  \in U_v}\mathbb{E}_{\rm{p}} ( z_{ij}^2\alpha_i^2) +\sum_{j=1}^{T_v} \dfrac{1}{N_v^2} \mathbb{E}_{\rm{p}}\left[\sum_{i\in U_v}\sum_{\ell\neq i  \in U_v}z_{ij} z_{\ell j}\mathbb{E}_{\rm{p}} ( \alpha_i \alpha_{\ell}|\mathcal A)\right].\label{Bv_tree}
		\end{eqnarray}
The first term on the right hand-side of (\ref{Bv_tree}) can be bounded as follows:
\begin{eqnarray}
\sum_{j=1}^{T_v} \dfrac{1}{N_v^2} \sum_{i  \in U_v} \mathbb{E}_{\rm{p}} (z_{ij}^2\alpha_i^2)  \leqslant \frac{1}{cN_v^2}\sum_{i\in U_v}\mathbb{E}_{\rm p}||\mathbf{z}_i||_2^2=\mathcal O\left(\frac{1}{n_v}\right), \label{Bv_tree_simple}
\end{eqnarray}	
where the $T_v$-dimensional vector $\mathbf{z}_i=(z_{ij})_{j=1}^{T_v}$ with $z_{ij}=\mathds{1}\left(\bx_i \in A_j\right)$ contains only one non-null value, so $||\mathbf{z}_i||_2=1$ for all $i\in U_v.$
Turning to the second on from the right-side  of (\ref{Bv_tree}), we have $|\mathbb{E}_p(\alpha_i \alpha_{\ell}\rvert \mathcal{A})|  \leqslant |(\pi_{i\ell}-\pi_i\pi_{\ell})/\pi_i \pi_{\ell}| + |R_{i\ell}|,$
where $R_{i\ell} = \dfrac{r_{i\ell} }{\pi_i \pi_{\ell}} - \dfrac{ r_{ii} }{ \pi_i} - \dfrac{ r_{\ell\ell}}{\pi_{\ell}}$. By Assumption (H\ref{H3}), there exists a positive quantity $C$ such that $|R_{i\ell}|\leq C \mbox{max}_{i,\ell\in U_v}|r_{i\ell}|.$ Then,
\begin{eqnarray}
&&\dfrac{1}{N_v^2} \mathbb{E}_{\rm{p}}\left[\sum_{j=1}^{T_v} \sum_{i\in U_v}\sum_{\ell\neq i  \in U_v}z_{ij} z_{\ell j}\mathbb{E}_{\rm{p}} ( \alpha_i \alpha_{\ell}|\mathcal A) \right] \nonumber\\
&\leqslant&  \dfrac{1}{N_v^2} \mathbb{E}_{\rm{p}}\sum_{j=1}^{T_v} \sum_{i\in U_v}\sum_{\ell\neq i  \in U_v}|z_{ij} z_{\ell j}||\mathbb{E}_{\rm{p}} ( \alpha_i \alpha_{\ell}|\mathcal A)|\nonumber\\
&\leqslant& \dfrac{1}{N_v^2}\sum_{j=1}^{T_v} \sum_{i\in U_v}\sum_{\ell\neq i  \in U_v}\left|\frac{\pi_{i\ell}-\pi_i\pi_{\ell}}{\pi_i \pi_{\ell}}\right|\mathbb{E}_{\rm{p}}(|z_{ij} z_{\ell j}|)+\dfrac{C}{N_v^2}\sum_{j=1}^{T_v}\mathbb{E}_{\rm{p}}\left[\mbox{max}_{i,\ell\in U_v}|r_{i\ell}| \sum_{i\in U_v}\sum_{\ell\neq i  \in U_v}|z_{ij} z_{\ell j}|\right]\nonumber\\
&\leqslant&\frac{n_v\max_{i,\ell\in U_v,i\neq \ell}|\pi_{i\ell}-\pi_i\pi_{\ell}|}{c^2n_vN_v^2} \mathbb{E}_{\rm{p}}\left(\sum_{j=1}^{T_v} \sum_{i\in U_v}\sum_{\ell\neq i  \in U_v}|z_{ij} z_{\ell j}|\right)\nonumber\\
&&+\dfrac{C}{N_v^2}\sum_{j=1}^{T_v}\mathbb{E}_{\rm{p}}\left(\mbox{max}_{i,\ell\in U_v}|r_{i\ell}| \sum_{i\in U_v}\sum_{\ell\neq i  \in U_v}|z_{ij} z_{\ell j}|\right)\nonumber\\
&\leqslant& \dfrac{n_v\max_{i,\ell\in U_v,i\neq \ell}|\pi_{i\ell}-\pi_i\pi_{\ell}|}{c^2n_vN_v}\sum_{i\in U_v}\mathbb{E}_{\rm{p}}||\mathbf{z}_i||_2^2+\frac{C}{N_v}\mathbb{E}_{\rm{p}}\left(\mbox{max}_{i,\ell\in U_v}|r_{i\ell}| \sum_{i\in U_v}||\mathbf{z}_i||_2^2\right)=\mathcal{O}\left(\dfrac{1}{n_{v}}\right)\nonumber\\
\label{Bv_tree_mixte}
\end{eqnarray}
by Assumptions (H\ref{H1})-(H\ref{H5}) and the fact that  $||\mathbf{z}_i||_2=1$ for all $i.$ From (\ref{Bv_tree}), (\ref{Bv_tree_simple}) and (\ref{Bv_tree_mixte}), it follows that 
\begin{eqnarray*}
\mathbb{E}_{\rm{p}} \left( \sum_{j=1}^{p_v}  b_j^2\right)=\mathcal{O}\left(\dfrac{1}{n_{v}}\right),
\end{eqnarray*}
which, combined with (\ref{borne_beta_tree}) in (\ref{terme_B_tree}), leads to
\begin{eqnarray}
\mathbb{E}_{\rm{p}}  \bigg\rvert \sum_{j=1}^{p_v} \widehat{\beta}_j b_j\bigg\rvert =\mathcal{O}\left(\sqrt{\frac{1}{n_{0v}}}\right).\label{borne_terme_B_tree}
\end{eqnarray}
Finally, from (\ref{decomp_tree}), (\ref{terme_A_tree}) and (\ref{borne_terme_B_tree}) we get that
\begin{eqnarray*}
\mathbb{E}_{\rm{p}} \bigg[ \dfrac{1}{N_v}\bigg\rvert \widehat{t}_{\rm{tree}} -t_y\bigg\rvert  \bigg] =\mathcal{O}\left(\sqrt{\dfrac{1}{n_{0v}}}\right).
\end{eqnarray*}

\end{document}